\documentclass[12pt, letterpaper]{article}

\usepackage[utf8]{inputenc}
\usepackage{amsmath}
\usepackage{mathrsfs}
\usepackage[table]{xcolor}
\usepackage{amsfonts}
\usepackage{amssymb}
\usepackage[margin=1in]{geometry}
\usepackage{graphicx}
\usepackage{subcaption}
\usepackage{enumerate}
\usepackage{color}
 \usepackage[colorlinks,citecolor=blue,urlcolor=blue,bookmarks=false,hypertexnames=true]{hyperref} 

 \usepackage{tikz}
 \usetikzlibrary{backgrounds}

\usepackage{cleveref}
\usepackage{pgfplots}
\usepackage{wasysym}
\usepackage[mathlines]{lineno}
\usepackage{comment}
\usepackage{setspace}
 \usepackage[semicolon]{natbib}
\usepackage{appendix}

\usepackage{todonotes}
\usepackage{lipsum}
\usepackage[ruled,linesnumbered]{algorithm2e}
\usepackage[normalem]{ulem}
\usepackage{csvsimple}
\usepackage{booktabs}

\newcommand{\maxRate}{R} 
\newcommand{\energyReq}{E}
\newcommand{\energyResCust}{E_i(t)}
\newcommand{\decVarCustTime}{r_{i}(k)}
\newcommand{\baseload}{\ell}
\newcommand{\minRate}{R^\mathrm{min}}
\newcommand{\activeCusts}{\mathcal{N}(t)}
\newcommand{\deadlineCustTime}{d_i(t)}

\newcommand{\deadline}{d}

\graphicspath{{Figures/}}

\title{ \bf 
Achieving Reliable Coordination of Residential Plug-in Electric Vehicle Charging: A Pilot Study
}
\author{Polina Alexeenko\thanks{Polina Alexeenko  and Eilyan Bitar are with the School of Electrical and Computer Engineering, Cornell University, Ithaca, NY, 14853, USA.  Emails: {\tt\small \{pa357, eyb5\}@cornell.edu}} \and Eilyan Bitar\footnotemark[1]
}

\date{}

\begin{document}

\begingroup
\nolinenumbers	
\maketitle
\endgroup

\begin{abstract}
We report findings from a real-world pilot study exploring a novel pricing and control mechanism to coordinate residential EV charging loads. The proposed pricing mechanism presents EV owners with a ``menu of deadlines’’ that offers lower electricity prices the longer they’re willing to delay their charging completion times. Given customers’ reported charging preferences, a smart charging system dynamically optimizes the power drawn by EVs in real-time to minimize their collective strain on the grid while ensuring all EVs are charged by their user-requested deadlines. We find that customers allow their charging to be delayed by over eight hours on average. Using this flexibility, the smart charging system reliably eliminates demand spikes by reshaping EV loads to flatten the aggregate load curve. Importantly, customer participation rates remained stable throughout the study, providing evidence that the proposed mechanism is a viable ``non-wires alternative’’ to meet the growing demand for electricity from EVs.
\end{abstract}

\section{Introduction} \label{sec:introduction}

The US transportation sector is on the brink of a major transformation. The internal combustion engine's hold on American transportation is beginning to slip, and for the first time a vision of an all-electric vehicle (EV) future is coming into focus.
Driven by declining battery costs \citep{nykvist2015rapidly, lutsey2019update,Hitting21Bloomberg}, progressive policy \citep{Fung16States, Dennis20California, outlook2020electric}, consumer demand \citep{outlook2020electric, daramy2019systematic, parker2021saves}, and the world's largest automakers \citep{colias_2021}, it's a vision that may become a reality in a matter of decades.  A recent study by the International Energy Agency anticipates that there will be between 140 and 245 million EVs on the road worldwide by 2030 \citep{outlook2020electric}.

But even with this great momentum, the transition to an  all-electric vehicle future won't be possible without careful coordination with the power grid.  If left unmanaged, the power demanded by many EVs charging at the same time in the evening will amplify existing peak loads \citep{quiros2018electric}, resulting in significant power losses \citep{clement2010impact}, reduced power quality  \citep{gruosso2016analysis, leou2013stochastic}, and potentially exceeding the grid's  capacity to reliably meet demand \citep{muratori2018impact, jardini2000distribution, gong2011study, hilshey2012estimating,powell2020controlled}.

To accommodate the increase in load driven by the unmanaged charging of EVs, electric power utilities and grid operators would need to build new generators to produce enough power, and expand transmission and distribution infrastructure to deliver that power to the electric vehicles. In states like Texas and California, where EV adoption is growing rapidly, these enhancements to the grid infrastructure would potentially cost tens of billions of dollars, and could take decades to complete \citep{Davidson18switching}.

Timing matters. While most EV owners typically begin charging their cars when they come home in the evening---when demand for electricity is peaking---their charging requirements are usually flexible in the sense that their EVs remained connected to their chargers long after they've completed charging. Providing concrete evidence for this claim,  \textit{The EV Project}---a nation-wide study spanning three years and involving more than 8000 EV owners---found that most EVs, when charging overnight, usually finished charging within three hours of plugging in, but remained connected to their chargers for an average of twelve hours \citep{smart2015lessons}.
In other words, most EV owners don't need their cars charged immediately, but within a reasonable window of time before they expect to unplug and  depart next for their next trip.
This finding suggests the possibility that  some EV owners, being flexible in this way, might be willing to delay the time required to charge their cars given the right incentive. 

 \subsection{Limitations of time-of-use pricing} \label{sec:limitations}
 
 In an effort to unlock this flexibility in EV charging, a number of electric power utilities have begun to offer their EV-owning residential customers time-of-use (TOU) rates, where the price of electricity varies over the course  of the day according to a predetermined and fixed schedule, typically being cheapest during off-peak hours. As of September 2019, there were over fifty different TOU rates being offered by  utilities to residential EV owners across the US \citep{residential2019sepa}.

While this might seem like a reasonable approach, nondiscriminatory\footnote{An electricity rate is defined to be \emph{nondiscriminatory} if all customers within a particular service class and territory are charged identical rates for their electricity consumption. Price discrimination is prohibited by the Federal Energy Regulatory Commission to limit the possibility of inequity among customers \citep{eisen2015ferc}.} TOU rates are constrained in terms of their ability to attenuate the impact that EVs will have on peak load.  By offering customers lower prices to charge during hours of the day that are ordinarily off-peak, TOU rates  can  have an unintentional synchronizing effect on EV charging patterns. For example, a number of utility-run trials have observed a ``timer peak" phenomenon in which many EV owners program their vehicles to begin charging simultaneously at the start of the off-peak pricing period \citep{smart2015lessons}. 
Related to this effect, several recent studies have shown that TOU rates can sharpen aggregate EV load profiles to such an extent that they accelerate distribution transformer aging more rapidly than unmanaged charging  \citep{wu2011load,hilshey2012estimating, powell2020controlled}. 
Importantly, these issues can arise even at relatively low levels of aggregate EV penetration, because EV registrations tend to be geographically clustered \citep{cluster2013inl}. 
TOU rates will ultimately fail to eliminate demand peaks driven by EV charging at scale. A greater degree of coordination is needed.

To address these drawbacks of TOU rate designs, a number of alternative dynamic pricing mechanisms have been proposed in the literature, where the retail price of electricity is allowed to vary in proportion to the realized aggregate demand over the course of the day \citep{ma2011decentralized, karfopoulos2012multi,tushar2012economics}. While such dynamic pricing schemes are  shown  to effectively `flatten' the aggregate demand profile in theory, their effectiveness hinges on the assumption that individual EV owners will possess the capacity to solve nontrivial decision making problems where they correctly anticipate the impact of their charging control strategies on the determination of prices. 
Another practical shortcoming of such dynamic pricing mechanisms is the risk they impose upon customers in the form of uncertainty about future prices and the subsequent payments that they must make.  Because income levels are known to be negatively correlated with risk aversion \citep{grable2016financial}, price uncertainty is likely to disproportionately impact lower income customers.

\subsection{Direct control of EV charging} \label{sec:direct}

Direct load control is an alternative approach to managed EV charging which addresses many of the shortcomings of dynamic pricing mechanisms while offering additional benefits such as the potential for vehicle-to-grid (V2G) services.
Under direct load control, a central coordinating authority (e.g., the utility) provides each participating customer with an incentive in exchange for the ability to directly control their EV charging subject to a set of constraints specified by the customer.
As a result,  the central coordinating authority can dynamically reshape EV charging profiles to optimize a variety of objectives including minimizing the cost of energy (e.g., \citep{jin2013optimizing, jiang2019real}), maximizing the integration of intermittent renewables (e.g. \citep{honarmand2014integrated, szinai2020reduced}), or minimizing peak aggregate load (e.g., \citep{gan2012optimal, zhang2017real}). 
There is a vast theoretical literature on direct control of electric vehicle charging; for example, Yang et al. \citep{yang2015computational} and Wang et al. 
 \citep{wang2016smart} survey a wide variety of approaches to scheduling EV charging load considering various objectives, computational methods, and models of vehicle and customer behavior.

While direct load control mechanisms have been studied extensively using theoretical models, there are relatively few real-world experimental studies to date evaluating their effectiveness.
To more effectively manage EV charging loads on their distribution networks, some utilities and charging facility operators have begun to explore the possibility of directly controlling EV chargers in both residential \citep{quiros2018electric, bauman2016residential} and workplace charging environments \citep{lee2016adaptive, chynoweth2014smart, bohn2016real, andersen2019parker}.  For example, in the ChargeTO Program \citep{bauman2016residential}, the Adaptive Charging Network Project \citep{lee2016adaptive}, and the ChargeForward program \citep{spencer2021evaluating},  participating EV owners  use a web or mobile interface to initiate a charging request---specifying when they need their EVs charged by---and a centralized smart-charging system actively manages the power being drawn by their EVs while ensuring that all EVs are fully charged by their user-specified departure times.  
An important drawback of these  frameworks for managed EV charging is that they do not provide their users with explicit incentives to accurately report their expected departure times. 
 As a result, users may underreport their departure times, preferring to have their EVs fully charged well in advance of their anticipated departure times. 
 Indeed, data from  the Adaptive Charging Network Project reveals that  users of their platform frequently underreport their departure times by more than several hours on average \citep{lee2019acndata}. 
 Such a reduction in  the charging flexibility provided by users constrains the extent to which their EV charging loads can be reshaped and shifted in time, which ultimately limits the effectiveness of managed EV charging programs and their capacity to minimize peak aggregate loads.

\subsection{The OptimizEV Project}
In this article, we report findings from the OptimizEV Project---a pilot study that tests a novel  rate structure and technology for managed  EV charging. As one of its primary objectives, OptimizEV seeks to maximize the charging flexibility procured from customers by offering them  monetary incentives to delay the time required to charge their EVs. Each time a customer initiates a charging session, they are presented with a ``menu of deadlines'' that offers lower electricity prices the longer they’re willing to delay the time required to charge their EV. 
Using their smartphones, customers  specify a desired state-of-charge and  select a corresponding deadline by which their requested energy must be delivered. Given a collection of active charging requests, a smart-charging system dynamically adjusts the power being drawn by  each participating EV in real time to minimize their collective impact on peak system load, while simultaneously ensuring that every customer’s car is charged up to the desired level by the requested deadline. Customers get their energy when they need it and the smart-charging system optimally coordinates the delivery of that energy to limit demand spikes.

In 2019, 34 plug-in EV owners residing in Tompkins County, New York were recruited  to participate in the OptimizEV Project on a voluntary first-come, first-served basis.\footnote{Appendix \ref{sec:demo}  provides demographic information that is representative of the participant population.}
A level-2 charging station with cellular communication and control capabilities  was installed in the home of each project participant. The project was divided into two phases: an initial two-month \emph{unmanaged charging phase} (Phase I)  where  baseline EV charging data was collected while  project participants were exposed to the preexisting flat electricity rate between January 1, 2020 -- February 29, 2020.  This was followed by a fifteen-month  \emph{managed charging phase} (Phase II), between  March 8, 2020 -- May 31, 2021, during which time all project participants were required to initiate charging sessions using the OptimizEV platform. Project participants were taught to use the OptimizEV platform during the transition week between Phase I and Phase II.

During Phase I of the OptimizEV Project, the unmanaged charging patterns of project participants frequently resulted in a substantial increase in peak aggregate load. The user charging data collected over the course of this project also sheds light on several undesirable effects that TOU pricing may have on EV load patterns.  We find that when many of the participating EV owners respond to the TOU price signal by  delaying their charging until the start of the off-peak pricing period, the synchronization of  EV power demand that results when many EVs start charging simultaneously  induces new aggregate demand peaks in the middle of the night that are sometimes sharper and larger in magnitude than the unmanaged aggregate demand peaks that  would have otherwise resulted in the absence of TOU rate-based incentives. This suggests that at increased EV penetration levels, passive load coordination mechanisms like TOU pricing will likely fail to attenuate the peak load associated with unmanaged charging, and may make matters worse.

During the managed charging phase of the project, the proposed incentive and control mechanism was  highly effective in shifting the majority of EV charging loads off-peak into the night-time valley of aggregate load curve. In particular, the increase in peak load driven by EV charging  was significantly reduced, if not entirely eliminated, on a majority of days during the fifteen-month span of the project. The observed efficacy of the mechanism is due in part to high customer participation rates that remained stable over the course of the pilot---customers frequently engaged in optimized charging sessions, allowing the smart charging system to delay the completion of their charging  by nine hours on average. Interestingly,  we also find that the majority of sessions in which users decide to opt out of controlled charging are typically characterized by inflexible charging requirements---revealing that users are less likely to opt out if they do in fact have flexibility to offer.  The sustained user engagement over the course of the pilot not only confirms the presence of substantial flexibility in real-world residential EV charging loads, but also demonstrates that EV owners are frequently  willing to cede control of this flexibility to a smart charging system given a modest monetary incentive.  Collectively, these observations provide concrete evidence in support of the proposed incentive and control mechanism as a potentially viable ``non-wires alternative'' to support the increased demand for electricity driven by the growing adoption of EVs.
\section{Flexibility-differentiated pricing  of EV charging services} \label{sec:price}

The underlying premise of the OptimizEV Project is that different customers have different needs when charging their electric vehicles. 
Some customers are flexible and may be  willing to delay the time required to charge their EVs in exchange for a discounted electricity price, while other less flexible customers will pay the full electricity price (or perhaps a premium on that price) to have their EVs charged as quickly as possible. Consequently, if customers are presented with a menu of charging options that are differentiated according to the length of time required to charge their EVs (with more flexible charging options being offered at lower  prices), then different customers will prefer to have their EVs charged with different degrees of flexibility.   
The flexibility procured from customers can, in turn, be utilized to optimally coordinate the charging patterns of EVs in a manner that minimizes the utility's supply and delivery  costs, while respecting customers' stated charging preferences. 
Ultimately, the ability to better align the utility's cost structure with the diverse preferences of customers, through this special form of product differentiation, has the potential to improve both supplier and customer welfare.\footnote{Of course, in order for such programs to be cost effective at scale, the benefit to the utility, in the form of reduced energy costs or avoided infrastructure costs, must exceed the  program implementation and operational costs. These include administrative costs, the cost of the rate discounts paid out to participating customers, and the cost of the additional metering, communication, and control equipment required to perform direct EV load control. }

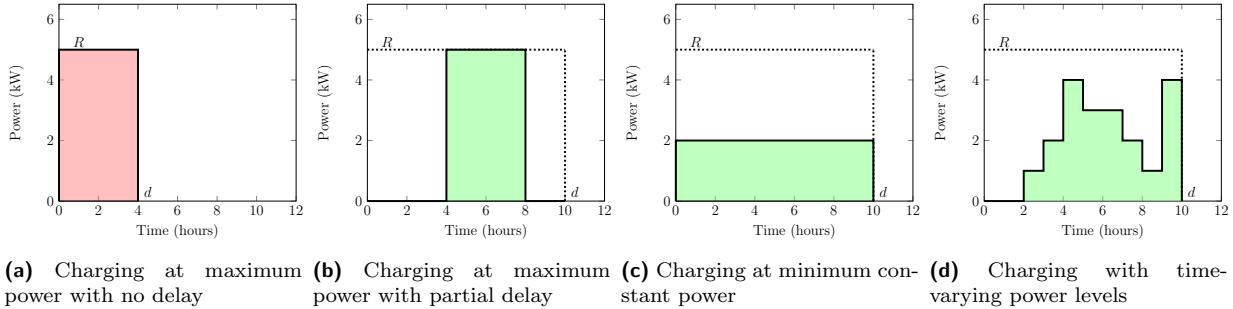
\begin{figure}[htpb] 
	\centering 
	\begin{minipage}{.24 \linewidth}
		\centering
		\label{fig:p3:c1} 
		\begin{tikzpicture}[scale=.46] 
		\begin{axis}[
		clip=false,
		axis line style=thick, 
		axis x line=box,
		axis y line=box,
		ymin=0,ymax=6.5, xmin=0,xmax=12, 
		xlabel=Time (hours), ylabel=Power (kW), grid=none,
		] 
		
		\coordinate (RLoc)  at (10, 525);
		\coordinate (DLoc) at (45, 30);

		\coordinate (Origin)   at (0,0);
		\coordinate (Bone) at (0,500);
		\coordinate (Btwo) at (40,500);
		\coordinate (Bthree) at (40,0);
		\draw [fill=red!50, fill opacity=0.5]    (Origin) -- (Bone) -- (Btwo) -- (Bthree) -- (Origin);  
		\draw [ultra thick, opacity=1] (RLoc) node {$\maxRate$}  (DLoc) node {$\deadline$}  (Origin) -- (Bone) -- (Btwo) -- (Bthree);
		
		\end{axis} 
		\end{tikzpicture} 
		\subcaption{Charging at maximum power with no delay} \label{fig:slackEX1}
	\end{minipage}
	\begin{minipage}{.24  \linewidth}
		\centering
		\begin{tikzpicture}[scale=.46] 
		\begin{axis}[
		clip=false,
		axis line style=thick, 
		axis x line=box, 
		axis y line=box, 
		ymin=0,ymax=6.5, xmin=0,xmax=12, 
		xlabel=Time (hours), ylabel=Power (kW),grid=none,
		] 
		\coordinate (RLoc)  at (10, 525);
		\coordinate (DLoc) at (105, 30);
		
		\draw[dotted, ultra thick] (RLoc) node {$\maxRate$} (0,500)  -- (100,500);
		\draw[dotted, ultra thick] (DLoc) node {$\deadline$} (100, 0)  -- (100,500);

		\coordinate (Origin)   at (0,0);
		\coordinate (Bzero) at (40,0);
		\coordinate (Bone) at (40,500);
		\coordinate (Btwo) at (80,500);
		\coordinate (Bthree) at (80,0);
		\coordinate (Bfour) at (100,0);
		\draw [fill=green!50, fill opacity=0.5] (Bzero) -- (Bone) -- (Btwo) -- (Bthree) -- (Bzero);  
		\draw [ultra thick] (Origin)-- (Bzero) -- (Bone) -- (Btwo) -- (Bthree) -- (Bfour); 
		
		\end{axis} 
		\end{tikzpicture} 
		\subcaption{Charging at maximum power with partial delay} \label{fig:slackEX2}
	\end{minipage}
	\begin{minipage}{.24 \linewidth}
		\centering
		\begin{tikzpicture}[scale=.46] 
		\begin{axis}[
		clip=false,
		axis line style=thick, 
		axis x line=box, 
		axis y line=box, 
		ymin=0,ymax=6.5, xmin=0,xmax=12, 
		xlabel=Time (hours), ylabel=Power (kW),grid=none,
		] 
		
		\coordinate (RLoc)  at (10, 525);
		\coordinate (DLoc) at (105, 30);
		
		\draw[dotted, ultra thick] (RLoc) node {$\maxRate$} (0,500)  -- (100,500);
		\draw[dotted, ultra thick] (DLoc) node {$\deadline$} (100,0)  -- (100,500);
		
		\coordinate (Origin)   at (0,0);
		\coordinate (Bone) at (0,200);
		\coordinate (Btwo) at (100,200);
		\coordinate (Bthree) at (100,0);
		\draw [fill=green!50, fill opacity=0.5] (Origin) -- (Bone) -- (Btwo) -- (Bthree) -- (Origin);  
		\draw [ultra thick] (Origin) -- (Bone) -- (Btwo) -- (Bthree); 
		
		\end{axis} 
		\end{tikzpicture}
		\subcaption{Charging at  minimum constant power }  \label{fig:slackEX3}
	\end{minipage}
	\begin{minipage}{.24  \linewidth}
		\centering
		\begin{tikzpicture}[scale=.46]  
		\begin{axis}[
		clip=false,
		axis line style=thick, 
		axis x line=box, 
		axis y line=box, 
		ymin=0,ymax=6.5, xmin=0,xmax=12, 
		xlabel=Time (hours), ylabel=Power (kW),grid=none,
		] 
		\coordinate (RLoc)  at (10, 525);
		\coordinate (DLoc) at (105, 30);
		
		\draw[dotted, ultra thick] (RLoc) node {$\maxRate$} (0,500)  -- (100,500);
		\draw[dotted, ultra thick] (DLoc) node {$\deadline$} (100,0)  -- (100,500);

		\draw[fill=green!50, fill opacity=0.5, draw=none] (20,0) -- (20,100) -- (30,100) -- (30,200)  -- (40, 200) -- (40,400) -- 
		(50,400) -- (50,300) -- (70, 300) -- (70, 200) -- (80,200) -- (80,100)  -- (90, 100) -- (90,400) -- (100, 400) -- (100, 0) -- (20,0);
		
		\draw [ultra thick] (0,0) -- (20,0) -- (20,100) -- (30,100) -- (30,200)  -- (40, 200) -- (40,400) -- 
		(50,400) -- (50,300) -- (70, 300) -- (70, 200) -- (80,200) -- (80,100)  -- (90, 100) -- (90,400) -- (100, 400) -- (100, 0);

		\end{axis} 
		\end{tikzpicture} 
		\subcaption{Charging with time-varying  power levels} \label{fig:slackEX4}
	\end{minipage}
	\caption{Four examples of feasible EV charging profiles associated with  (a) an inflexible charging request and (b)-(d) a flexible charging request. (a) EV Charging profile induced by  an inflexible charging request given by ($\maxRate = 5$ kW, $\energyReq = 20$ kWh, $\deadline = 4$ hours).  
		Because this charging request results in a slack time of $s = 0$ hours, the EV is charged without delay  at its maximum rate until its deadline. (b)-(d)  Three different EV charging profiles that satisfy a flexible charging request given by  ($\maxRate = 5$ kW, $\energyReq = 20$ kWh, $\deadline = 10$ hours). This charging request has a slack time of $s=6$ hours.}
\end{figure}

As part of the OptimizEV Project, we price differentiate customers'  charging requests according to a standard  measure of scheduling flexibility  in queueing systems known as \emph{slack time}. 
The slack time associated with a charging request is defined  as the length of time between the minimum time required to supply the desired state-of-charge (when charging at  the maximum rated power) and the maximum time that the customer is willing to wait to receive that energy. 
More formally, the \emph{slack time}  $s$ (units: hours) of  an EV charging request is a function of three user-specified parameters:  (1) the amount of \emph{energy} $\energyReq$ (units: kilowatt-hours) that the customer would like to receive;  (2) the maximum amount of time that the customer is willing to wait to receive their requested energy, which we referred to as their \textit{deadline} $\deadline$ (units: hours);  and   (3) the \emph{maximum rate}  $\maxRate$ (units: kilowatts) at which the customer's EV can be charged.  It follows that the slack time associated with a charging request can be  expressed as
\begin{linenomath*}
\begin{align} \label{eq:slack_def}
s = \deadline - \frac{\energyReq}{\maxRate},
\end{align}
\end{linenomath*}
where $\energyReq/\maxRate$ equals the minimum amount of time required to supply $E$ kilowatt-hours to an EV at a maximum charging power of $R$ kilowatts. Equation \eqref{eq:slack_def} offers an intuitive interpretation of  slack time as the extra amount of time  available to complete a charging request.   It follows that a charging request with zero slack ($s=0$) is completely inflexible, since the corresponding EV must be charged at maximum power from  start to finish  in order to deliver the customer's desired energy  by their deadline. Figure \ref{fig:slackEX1}  depicts an inflexible charging profile induced  by a charging request with zero slack time.
By contrast, a charging request with positive slack ($ s >0$) can be satisfied using one of a wide variety of charging profiles, some of which are illustrated in Figures \ref{fig:slackEX2}-\ref{fig:slackEX4}.  The greater the slack time in a charging request, the greater its flexibility.

In exchange for their willingness to delay the time required to charge their EVs, customers receive a  discount on the total cost of electricity required to charge their EVs given by
\begin{linenomath*}
\begin{align} \label{eq:charging_discount_formula}
\text{EV Charging Discount} =  p(s) \times \energyReq.
\end{align}
\end{linenomath*}
Here,  $s$ denotes  the slack time associated with a particular customer's charging request ($\energyReq$, $\deadline$, $\maxRate$), and $p(s)$ (units: \$/kilowatt-hour) denotes the corresponding discount (per unit of requested energy) given to that customer. To incentivize flexibility in the charging requests made by customers, the proposed \emph{slack-differentiated discount mechanism}   is structured so that (i) completely inflexible charging requests with zero slack receive no discount, i.e., $p(0) = 0$;  and (ii) charging requests with  larger slack times are rewarded with larger discounts, i.e.,  $p(s) \geq p(s')$ for all $s \geq  s'$.  We also note that customers are not permitted to make charging requests with negative slack ($s < 0$), since such requests are impossible to satisfy by their deadlines. 

\begin{figure}[htpb] 
		\centering
		\begin{tikzpicture}[scale=.8] 
		\begin{axis}[
		height=2.5in,
		width=4.5in, 
		clip=false,
		axis line style=thick, 
		axis x line=box,
		axis y line=box,
		ymin=0,ymax=.06, xmin=0,xmax=14, 
		xlabel= Slack (hours), ylabel= $p(s)$ \, (\$/kWh), grid=none,
		] 
		\draw [ultra thick, opacity=1] (20, 475) node {$p^{\rm max}$}   (0,0) -- (100,430) -- (140,430);
		\draw[dotted, ultra thick] (110, 45) node {$s^{\max}$}   ( 0,430)  -- (100,430)  -- (100,0);						
		\end{axis} 
		\end{tikzpicture} 
		\caption{A graphical illustration of the slack-differentiated discount mechanism used in the OptimizEV Project. The discount  $p(s)$ (per unit of request energy) given to customers  increases linearly from $p(0) = 0$ $\cent$/kWh to the maximum available discount  of $p^{\rm max} = 4.3$ $\cent$/kWh   at a maximum  slack time of    $s^{\rm max} = 10$ hours. While customers are permitted to make charging requests with slack times exceeding  $s^{\rm max}$, they are not rewarded for providing additional flexibility beyond this ten hour threshold.}
\label{fig:price_mech} 
\end{figure}
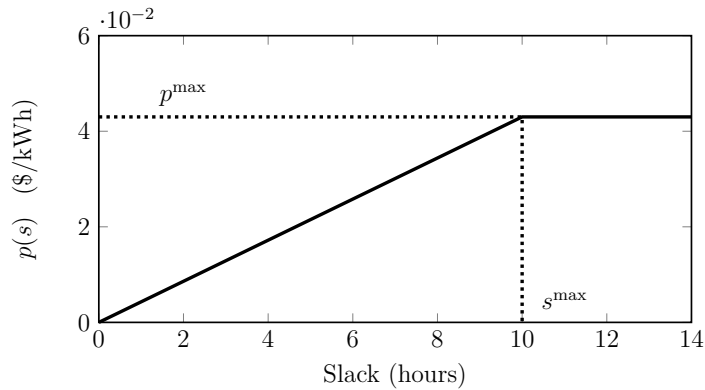 

Apart from these requirements, the structure of the proposed discount mechanism is highly versatile and can be adapted to reflect the  value of charging flexibility (slack time) as determined by the particular cost structure being optimized by the utility. 
For example, the discounts can be chosen to reflect  expected distribution-level infrastructure costs that can be avoided through optimized charging using the  flexibility provided by customers. In the OptimizEV Project, specifically, the slack-differentiated discount mechanism $p(s)$ is structured as a piecewise-linear function of the  slack time $s$  (depicted in Figure \ref{fig:price_mech}). The maximum discount (per unit of requested energy) available to customers under this mechanism was chosen to equal the delivery rate paid by residential electricity customers in NYSEG's territory at the time of the pilot.\footnote{An electric utility's delivery rate reflects costs associated with the infrastructure required to transport and distribute energy (e.g., transformers, feeders, and substation hardware).}

\section{Optimized delivery of flexible EV charging services} \label{sec:sch}

Given a collection of active charging requests,  an effective smart-charging system must coordinate the charging patterns of the corresponding  EVs to minimize their aggregate contribution to peak system load, while ensuring that every EV is charged by its requested deadline. This coordination problem is complicated by the presence of many sources of uncertainty, including uncertainty in the timing and nature of future charging requests,  unexpected variations in total system load,  fluctuating wholesale electricity prices, unanticipated EV charging characteristics, and failures in the communication network being used for control.

\begin{figure}[h!]
	\centering
	\begin{minipage}{.3 \linewidth}
		\centering
	     \includegraphics[width=.95\linewidth]{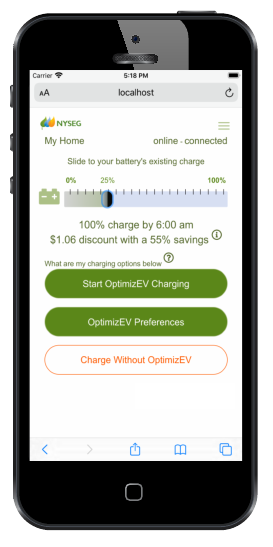}
		\subcaption{Landing page}
		\label{fig:ui_landing}
	\end{minipage}
	\begin{minipage}{.3 \linewidth}
		\centering
		\includegraphics[width=.95 \linewidth]{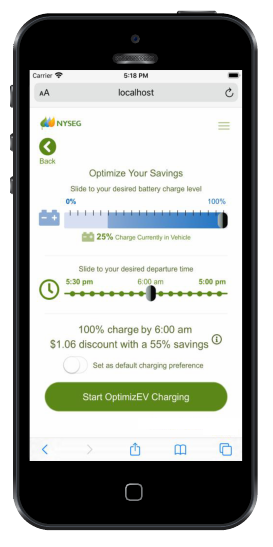}
		\subcaption{Charging preferences page}
		\label{fig:ui_selection}
	\end{minipage}
	\begin{minipage}{.3 \linewidth}
		\centering
		\includegraphics[width=.95\linewidth]{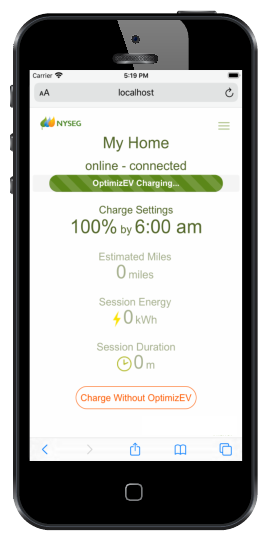}
		\subcaption{Charging session status page}
		\label{fig:ui_status}
	\end{minipage}
	\caption{User interface of the OptimizEV system. Subfigure (a) shows the landing page, where customers select whether to engage in a controlled or uncontrolled charging session. Subfigure (b) shows the preferences page, where customers select their controlled charging session preferences. Subfigure (c) shows the status page, where customers can see information about an ongoing charging session (e.g., which preferences were selected, the session's duration, and how much energy has been delivered).} 
	\label{fig:ui}
\end{figure}

In order to effectively adapt to changing system conditions in real time, the smart-charging system developed for the OptimizEV Project utilizes a model predictive control approach \citep{camacho2013model} to continuously re-optimize the active EVs' charging profiles every  minute of the day. At a high level, the OptimizEV smart-charging system functions as follows:

\begin{enumerate}
\item \emph{Customer Inputs:} To initiate a charging session, a customer begins by connecting their EV to their charging station and logging into the OptimizEV user interface (UI) using their smart phone. The UI (depicted in Figure \ref{fig:ui}) allows customers to make charging requests and monitor the progress of their ongoing charging sessions. When making a charging request, a customer must provide their EV's present state-of-charge (SOC),  in addition to providing their   desired state-of-charge and completion deadline. The difference between their desired SOC and present SOC is used to calculate their energy requirement. Customers can  also choose to opt out of  optimized charging at any time prior to or during an ongoing charging session. To limit their ``interaction costs'' with the UI, customers can also preset default charging preferences, bypassing the need to manually input these preferences in subsequent charging sessions.

\item \emph{Data Acquisition:}  Every minute, the smart-charging system collects information from all  newly  and previously connected EVs that are actively being charged. These measurements include connection status, current state-of-charge, energy consumption, and the actual power drawn by active EVs during the previous one-minute time interval.

\item  \emph{Computation:}  This information is passed to an optimization algorithm, which determines optimal charging profiles for all active EVs. These optimal charging profiles are designed to collectively minimize  peak aggregate load, while respecting the individual constraints associated with each EV, e.g., charging completion deadlines, charging rate constraints, and battery charging dynamics. 

\item \emph{Control:}   The optimized charging profiles are then transmitted to each EV's charging station as a sequence of time-varying power commands that each EV is instructed to track. The EVs adjust their  charging rates to track the updated power commands as closely as possible, while respecting the safe operating limits specified by their battery management systems. We use the SAE J1772 standard for  level-2 charging stations to dynamically control  each vehicle's charging rate.

\item \emph{Repeat:}   This data acquisition, computation, and control process repeats periodically every minute to enable real-time adaptation to changing system conditions and unexpected EV charging characteristics, while ensuring the complete satisfaction of all customers' charging requests and constraints.

\end{enumerate}
We refer the reader to Appendix \ref{sec:schDetail} for a detailed description of the underlying optimization model and the model predictive control  algorithm used to repeatedly optimize the EV charging profiles in the manner described above. In Appendix \ref{sec:practical}, we discuss a number of practical challenges encountered when deploying the proposed real-time scheduling algorithm in a real-world setting.

\begin{figure}[htb!]
\centering
	\begin{minipage}[b]{.75\linewidth}
	\centering
	\includegraphics[width=1\linewidth, trim=3cm 12.5cm 1cm 1cm,clip]{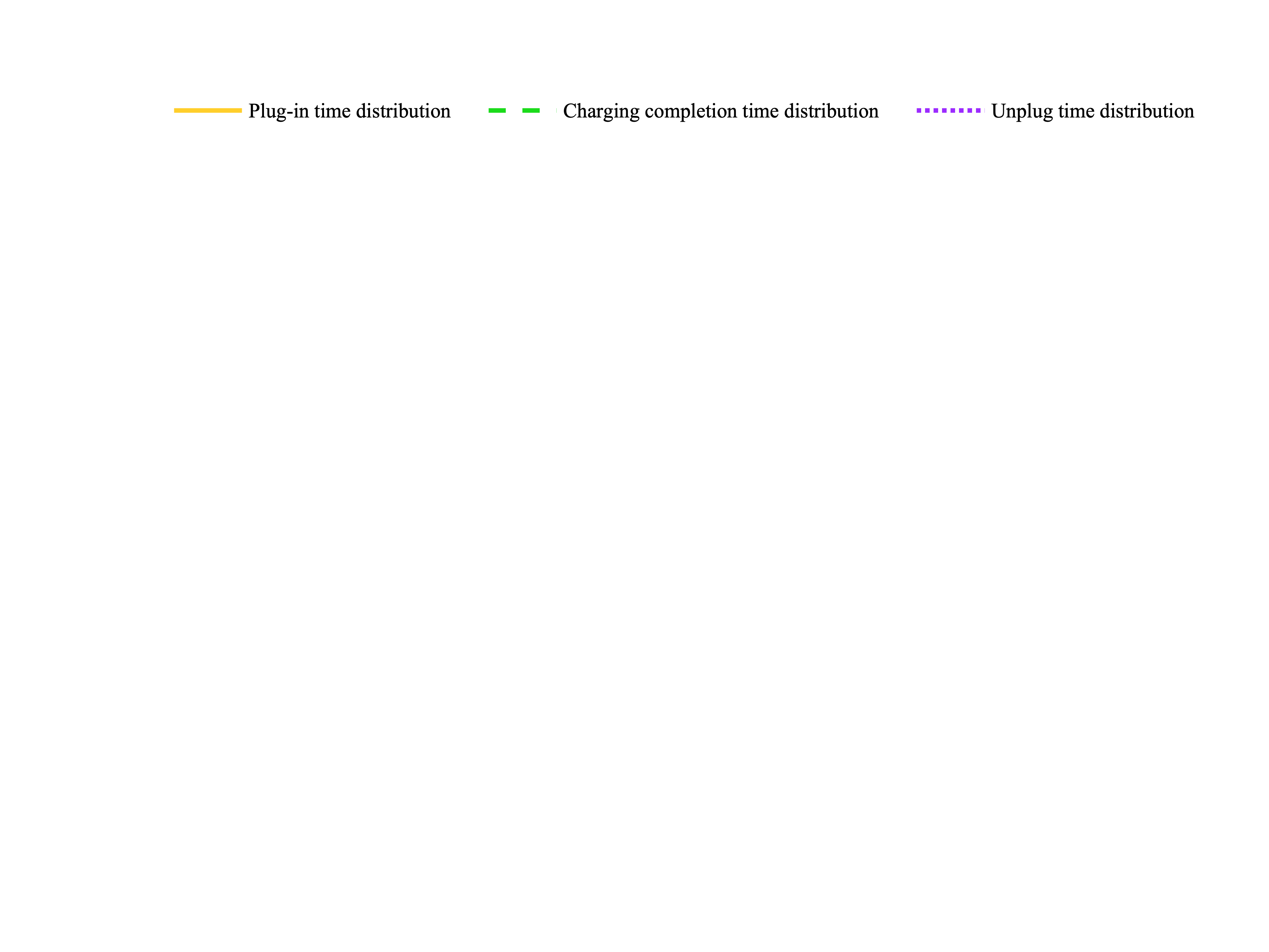}
\end{minipage}

\begin{minipage}[b]{.47\linewidth}
	\includegraphics[width=1\linewidth]{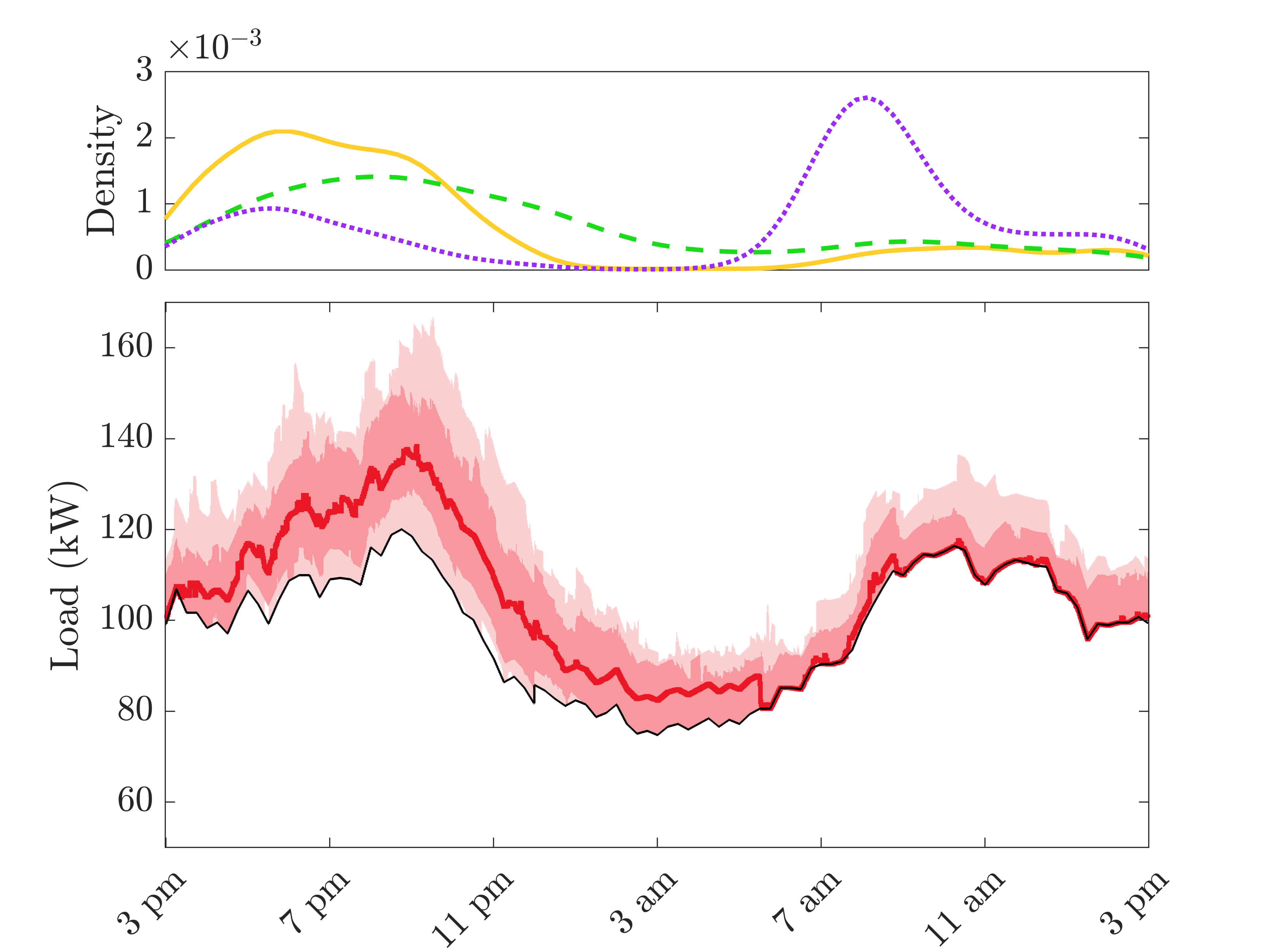}
	\subcaption{Weekday unmanaged charging patterns}
	\label{fig:loadTimeWD}
\end{minipage}
\begin{minipage}[b]{.47\linewidth}
	\includegraphics[width=1\linewidth]{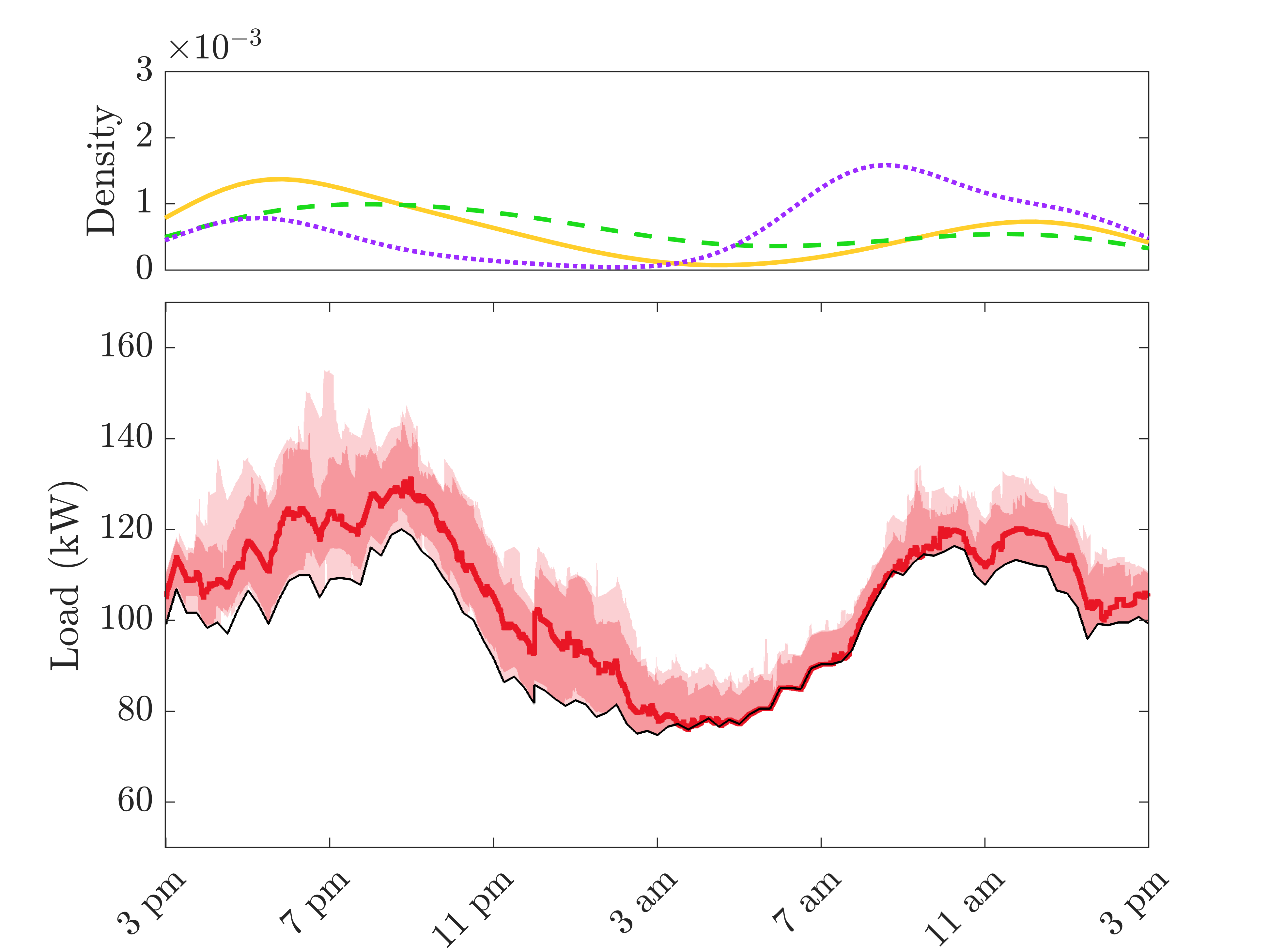}
	\subcaption{Weekend unmanaged charging patterns}
	\label{fig:loadTimeWE}
\end{minipage}

\caption{Unmanaged charging patterns during Phase I (January 1, 2020 to February 29, 2020) of the OpimizEV Project for weekdays (left column) and weekends (right column). The top figures depict kernel density estimates of session start times (solid yellow), session end times (dotted purple), and the times at which charging is finished (dashed green). The bottom figures provide plots of aggregate unmanaged EV charging patterns during Phase I.
The baseline (non-EV) load is depicted as a solid black curve. The median EV charging load is shown as a solid red curve, the interdecile range of the EV charging load is shaded in dark red, and the range between the maximum and minimum EV charging load is shaded in light red.} 
\label{fig:loadsTimesDistribution}
\end{figure}
\section{Unmanaged charging patterns and latent flexibility} \label{sec:phaseI}
During Phase I of the OptimizEV Project (January 1, 2020 to February 29, 2020), all customers participated in unmanaged charging: their EVs were charged at maximum power whenever connected to their in-home charging stations, and charging was not delayed or controlled. During this time, we continuously monitored and recorded the connection status and  power drawn by each user's EV at  regular one-minute intervals. In our analysis, we make use of EV connection and disconnection times as surrogates for user arrival and departure times to and from their homes. In Figure \ref{fig:loadsTimesDistribution}, we summarize several key attributes of unmanaged charging patterns revealed by these data.

Using the  data collected during the unmanaged charging phase of the project, we are able to empirically estimate distributions over user plug-in times, unplug times, and charging completion times, which are depicted in the upper plots of Figures \ref{fig:loadTimeWD}-\ref{fig:loadTimeWE}. Unsurprisingly, these distributions reveal that users typically plug in and start charging in the late afternoon and early evening, and typically unplug the following morning, presumably departing for their next trip. The resulting concentration of unmanaged EV charging loads in the evening is shown to manifest in a substantial amplification of peak load. The maximum and median increases in weekday peak loads are observed to be 47 kW (39\%  of the baseline peak) and 18 kW (15\% of the baseline peak), respectively. The impact of unmanaged EV charging patterns on peak load is less pronounced on weekends, as user plug-in times are more dispersed, which results in a more even spreading of EV charging loads across time.\footnote{The baseline (non-EV) load profile used in our analysis is based on load data taken from a primarily residential distribution circuit in Tompkins County, NY, which we re-scaled to reflect an aggregate demand profile associated with approximately 60 households. We refer the reader to  Appendix \ref{sec:baseline} for a more detailed description of how the baseline load profile is constructed.}

\begin{figure}[htb!]
\centering
	\includegraphics[width=1\linewidth, trim= 1cm 4.5cm 1cm 0cm ]{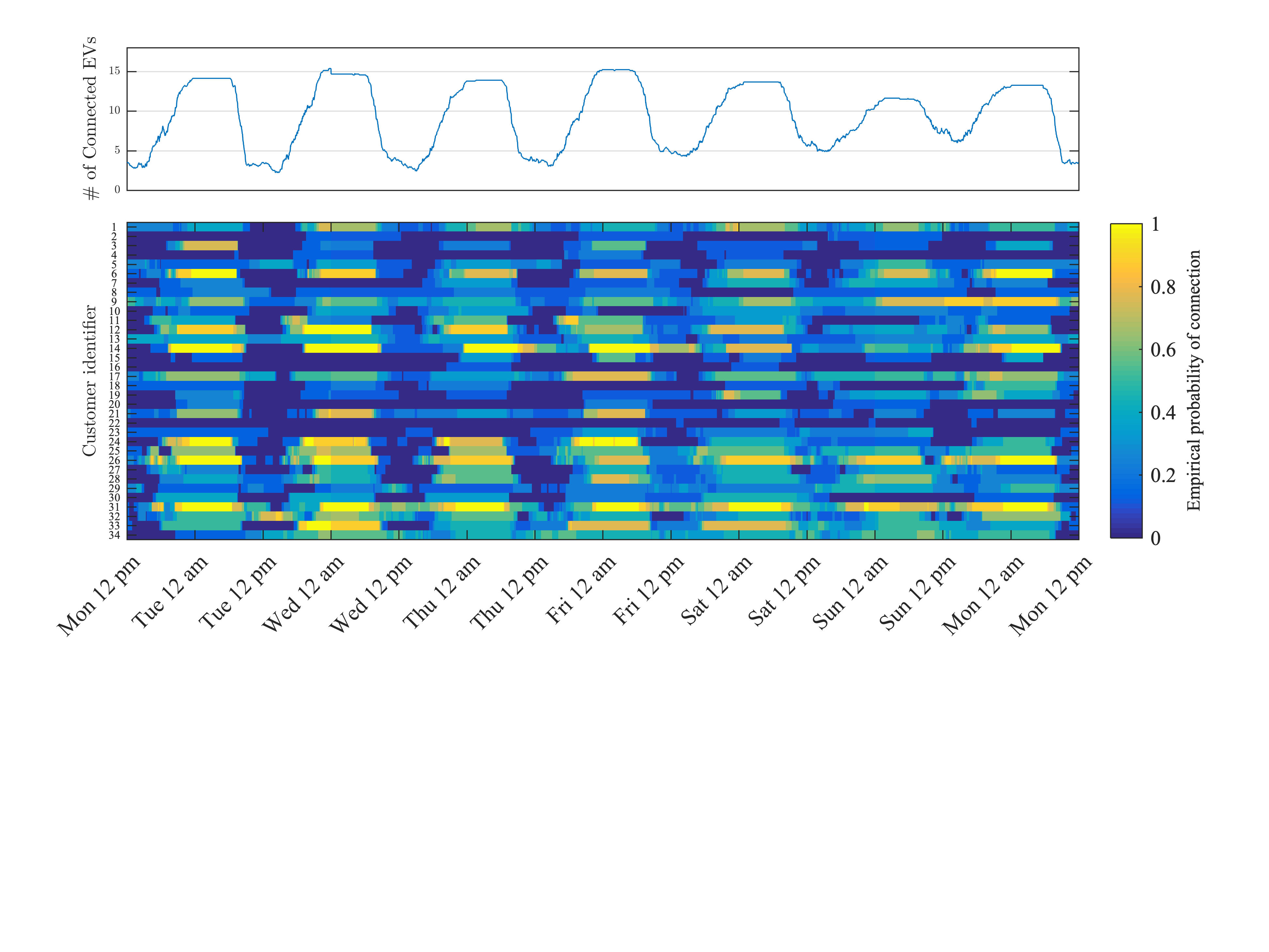}
\caption{User-level connection patterns during Phase I (January 1, 2020 to February 29, 2020) of the OptimizEV Project.  Each row of the heatmap depicts the empirical probability that a particular user's vehicle is connected during each minute of the week. The plot above the heatmap depicts the average, lower quartile, and upper quartile of the number of EVs connected to the grid throughout the week.}
\label{fig:connec}
\end{figure}

We also note that, while there is significant heterogeneity in the day-to-day connection patterns across different users, the aggregate behavior of users is more predictable. As can be seen in Figure \ref{fig:connec},  the  total  number of  EVs connected to the grid as a function of time exhibits a clear diurnal pattern, with the total number of EVs connected overnight averaging between 14-15 vehicles on weekdays and 11-13 vehicles on weekends.

The unmanaged charging data also reveals that users typically keep their EVs connected to their charging stations far longer than the amount of time required to fully charge their batteries. In particular, we find that EVs remained plugged in for 11 hours on average, while actively drawing power for only 2 hours on average. This implies that charging requests have an average slack time of 9 hours---revealing the presence of a significant amount of latent flexibility in users' charging requirements, and the potential to harness this flexibility to  minimize the contribution of residential EV charging loads to evening peak demand.

	\begin{figure}[h!]
	\centering
	\begin{minipage}[b]{.5\linewidth}
		
		\end{minipage}
	\begin{minipage}[b]{.65\linewidth}
	\centering
	\includegraphics[width=1\linewidth, trim=3cm 12cm 0cm 1.25cm,clip]{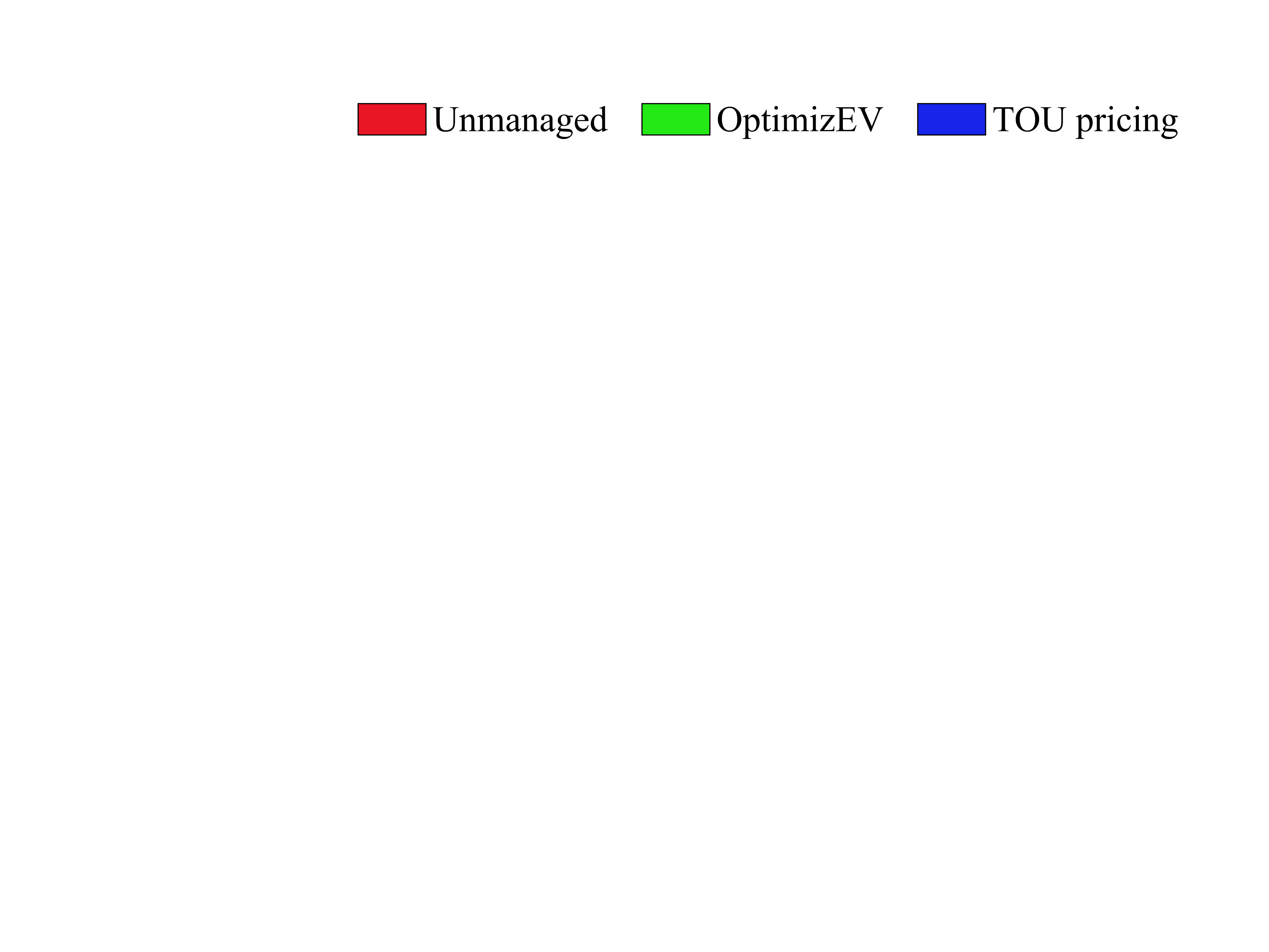}
	\end{minipage}
		\begin{minipage}[b]{.95 \linewidth}
		\centering
		\includegraphics[width=1\linewidth, trim=0cm 0cm 0cm -.5cm,clip]{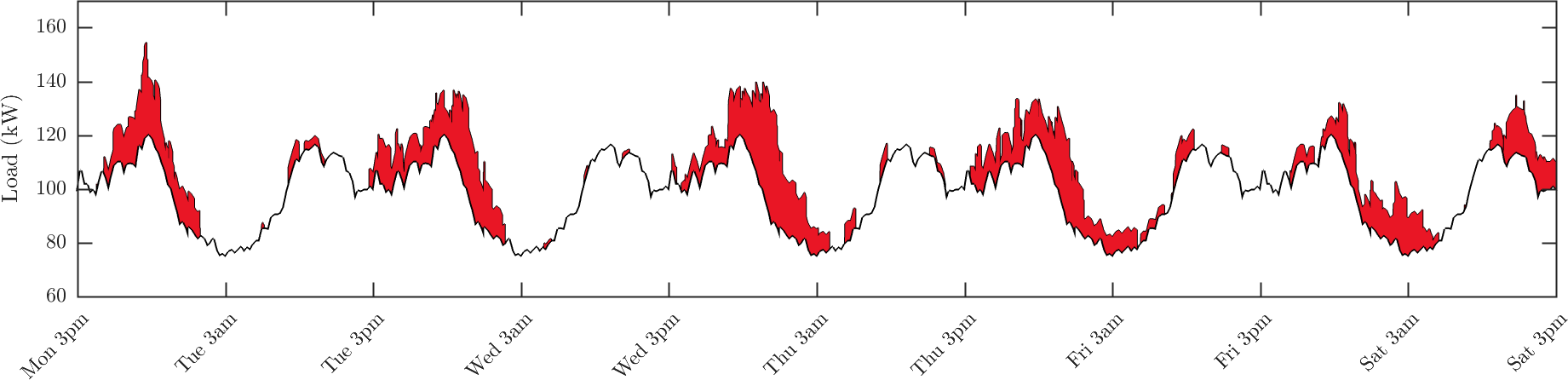}
		\subcaption{Unmanaged}
		\label{fig:weekLoadUnc}
	\end{minipage}
	\begin{minipage}[b]{.95 \linewidth}
		\centering
		\includegraphics[width=1\linewidth, trim=0cm 0cm 0cm -.5cm,clip ]{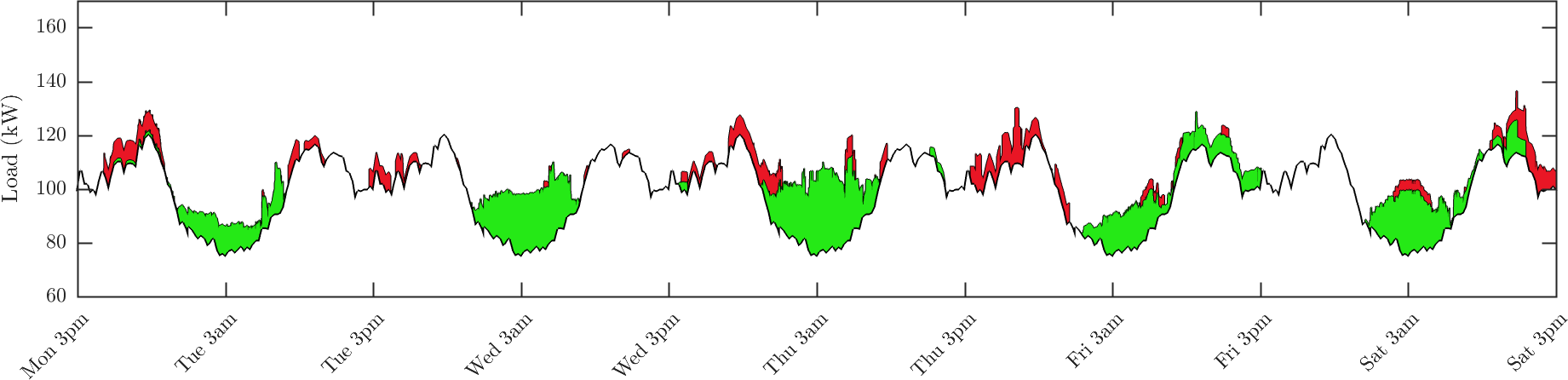}
		\subcaption{OptimizEV}
		\label{fig:weekLoadOpt}
	\end{minipage}
	\begin{minipage}[b]{.95 \linewidth}
		\centering
		\includegraphics[width=1\linewidth, trim=0cm 0cm 0cm -.5cm,clip]{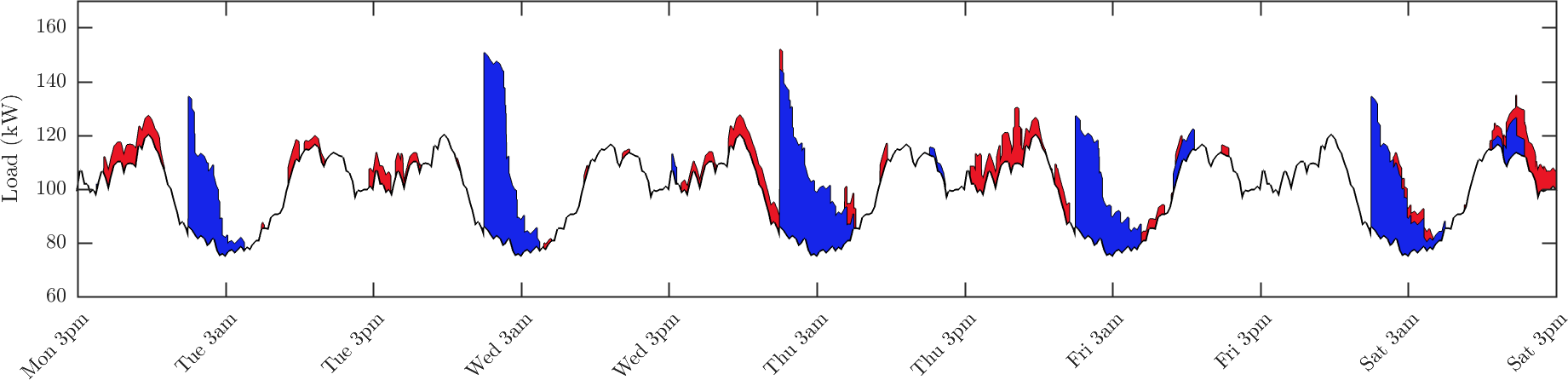}
		\subcaption{TOU pricing}
		\label{fig:weekLoadTOU}
	\end{minipage}

	\caption{Aggregate load profiles  (a) simulated under unmanaged charging, (b) realized under managed (OptimizEV) charging,  and (c) simulated under TOU pricing  between March 9, 2020 and March 14, 2020. In each subfigure, the baseline load is depicted by a black curve, unmanaged EV loads are shown in red, managed (OptimizEV) loads are shown in green, and loads under TOU pricing are shown in blue.}
	\label{fig:weekLoads}
\end{figure}

\clearpage

\begin{figure}[htb!]
	\centering
		\begin{minipage}{.32 \linewidth}
		\centering
		\includegraphics[width=1\linewidth, trim=0cm 0cm 1.5cm 0cm,clip]{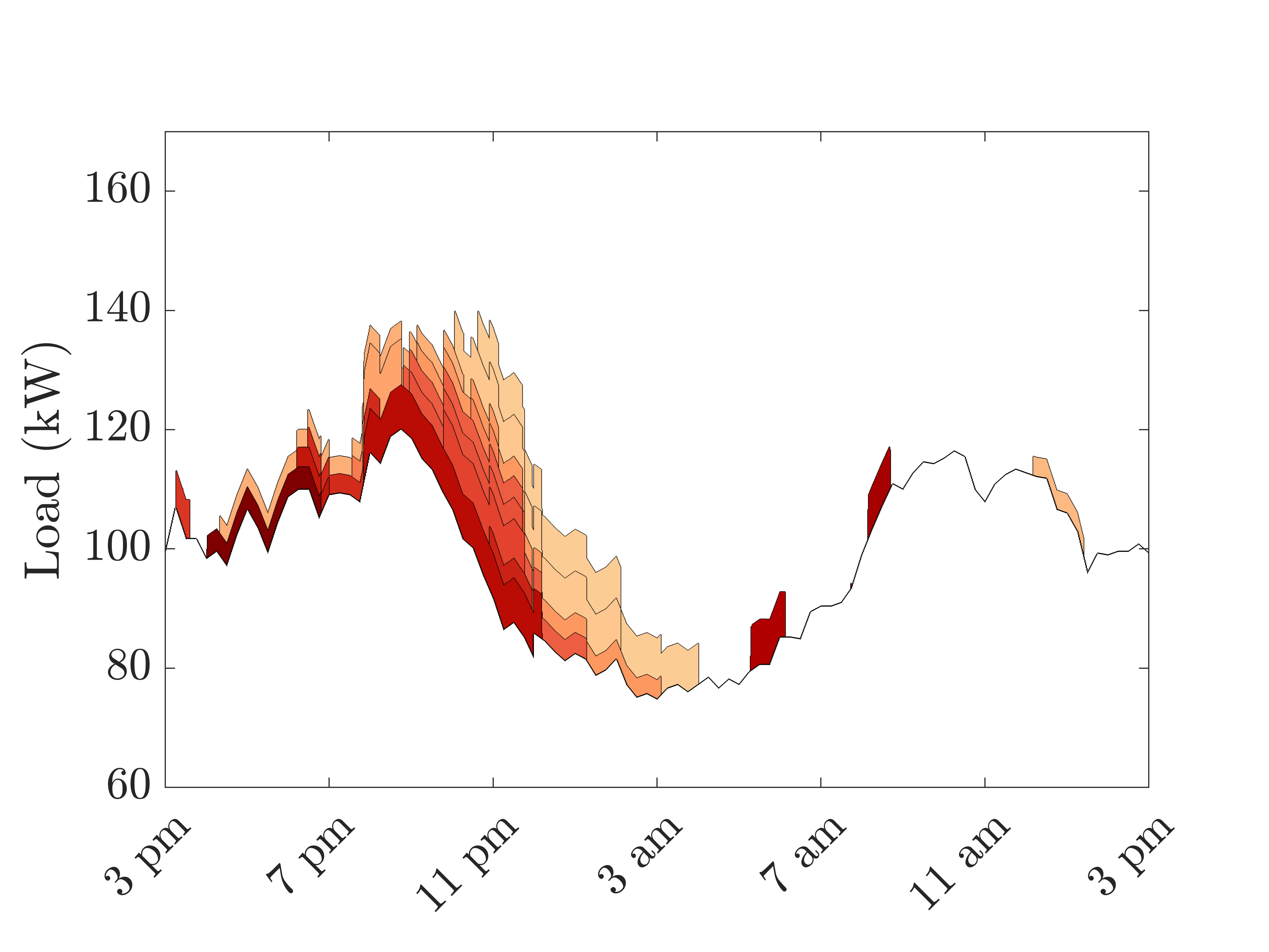}
		\subcaption{Unmanaged}
		\label{fig:dayLoadsUnc}
	\end{minipage}
	\begin{minipage}{.32 \linewidth}
		\centering
		\includegraphics[width=1\linewidth,  trim=0cm 0cm 1.5cm 0cm,clip]{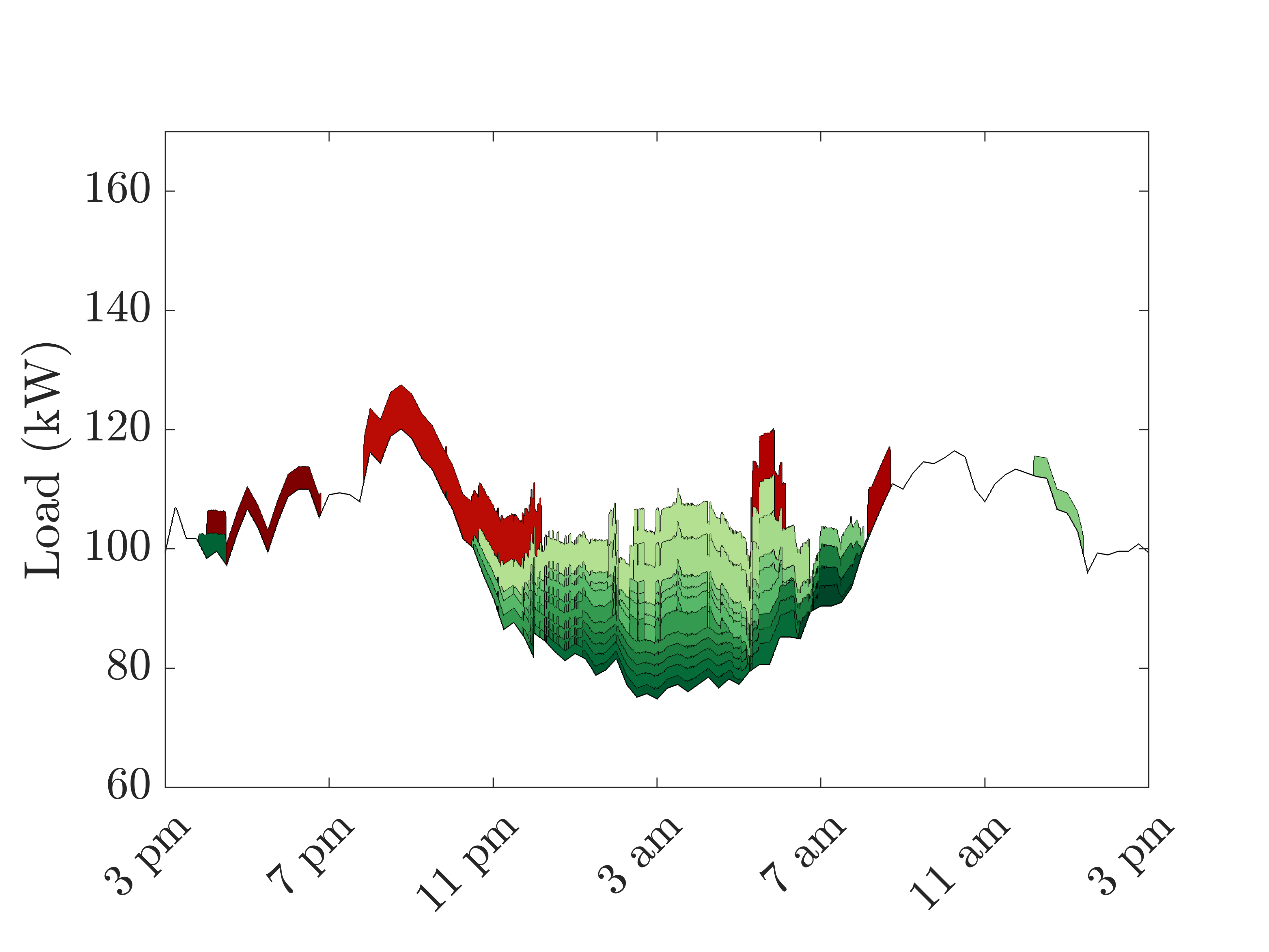}
		\subcaption{OptimizEV}
		\label{fig:dayLoadsOpt}
	\end{minipage}
	\begin{minipage}{.32 \linewidth}
		\centering
		\includegraphics[width=1\linewidth,  trim=0cm 0cm 1.5cm 0cm,clip]{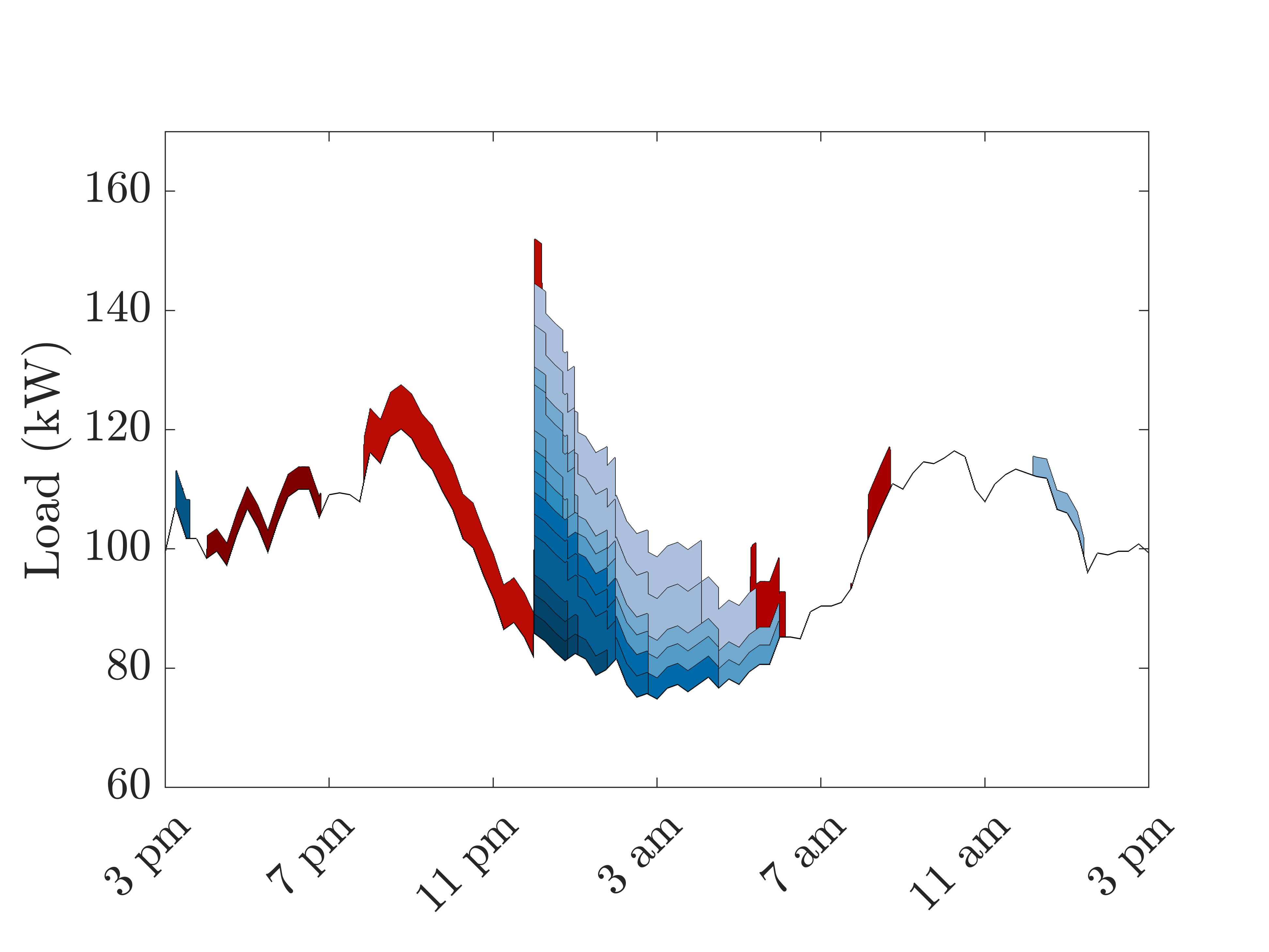}
		\subcaption{TOU pricing}
		\label{fig:dayLoadsTOU}
	\end{minipage}
	
	\begin{minipage}[b]{.32 \linewidth}
		\includegraphics[width=1\linewidth,  trim=0cm 0cm 1.5cm 0cm,clip]{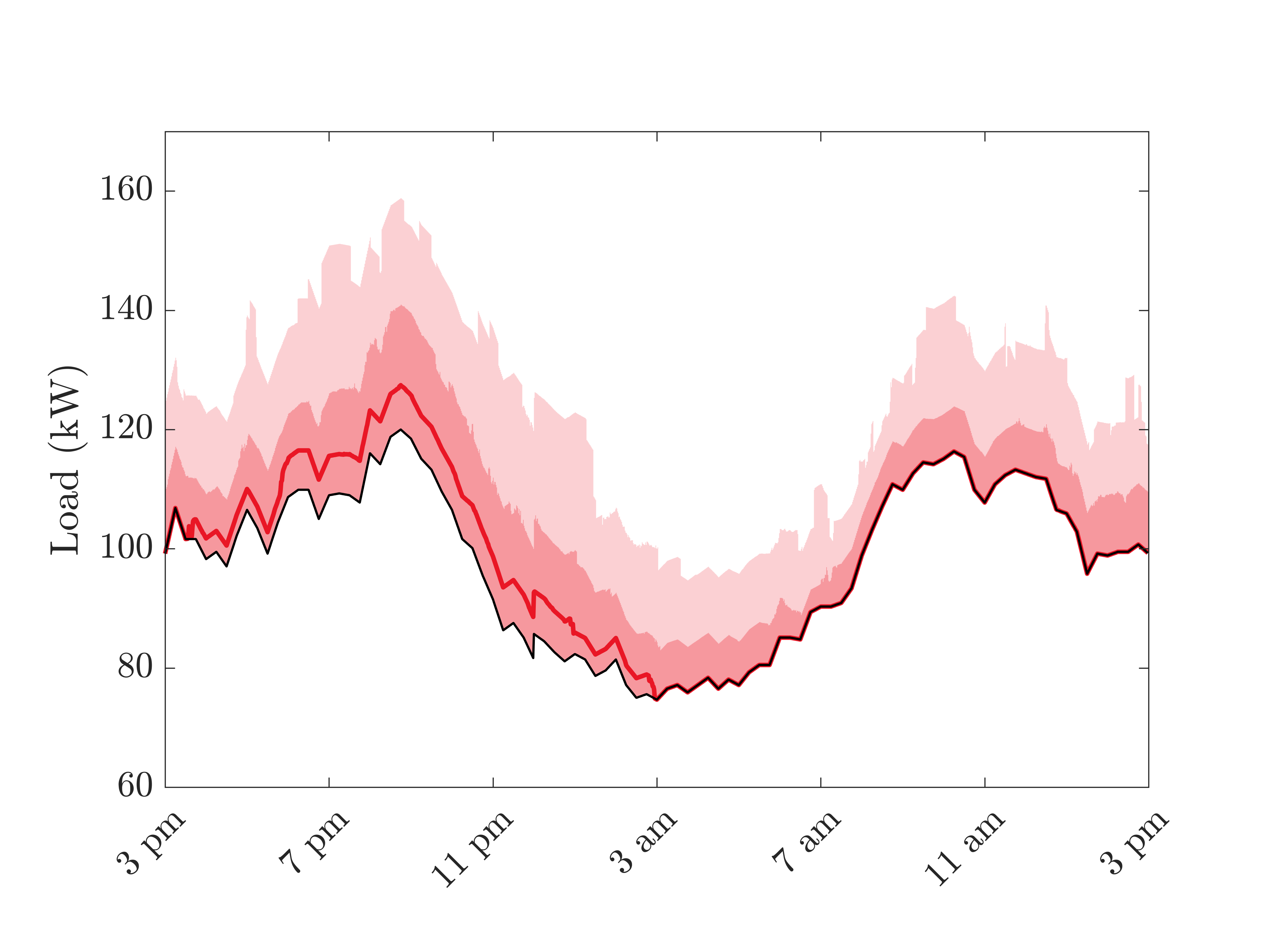}
		\subcaption{Unmanaged}
		\label{fig:dailyUncConf}
	\end{minipage}
	\begin{minipage}[b]{.32 \linewidth}
		\centering
		\includegraphics[width=1\linewidth, trim=0cm 0cm 1.5cm 0cm,clip]{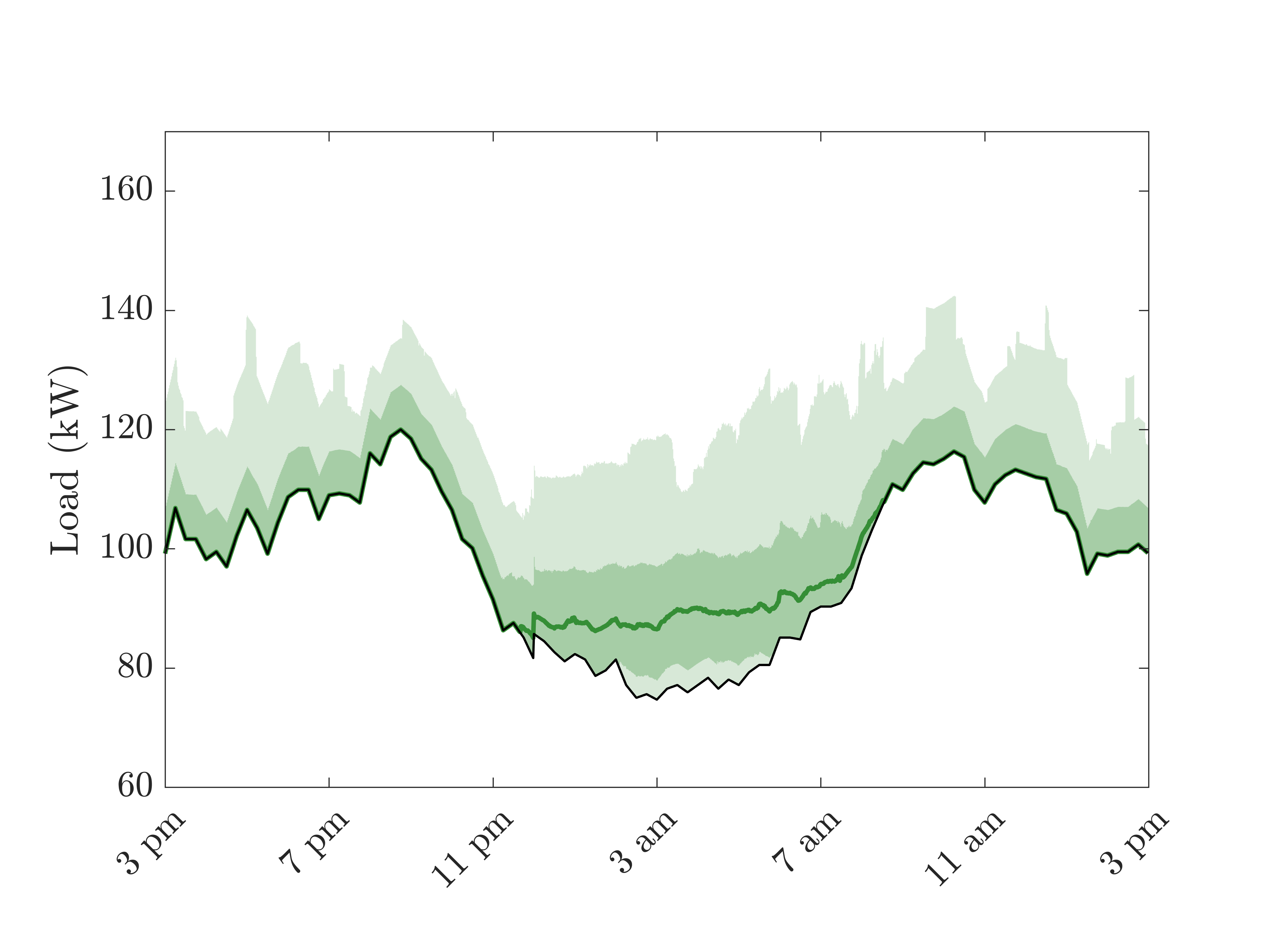}
		\subcaption{OptimizEV}
		\label{fig:dailyRealizedConf}
	\end{minipage}
	\begin{minipage}[b]{.32 \linewidth}
		\includegraphics[width=1\linewidth,  trim=0cm 0cm 1.5cm 0cm,clip]{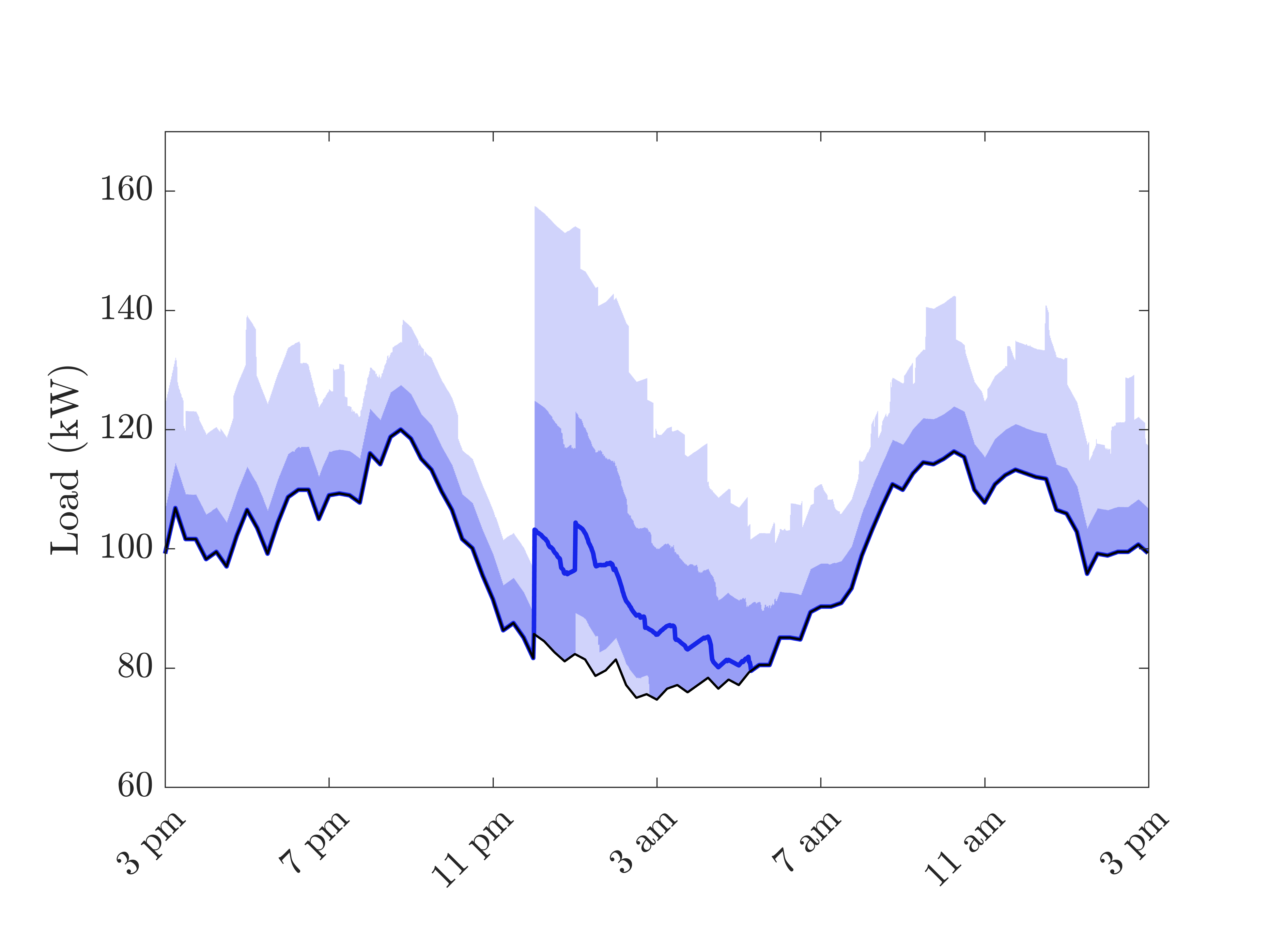}
		\subcaption{TOU pricing}
		\label{fig:dailyTOUConf}
	\end{minipage}

	\caption{ Daily charging patterns under unmanaged charging (left column), OptimizEV (middle column), and TOU pricing (right column). Subfigures (a)-(c) depict aggregate loads  realized under each scenario on March 11, 2020. In each subfigure, the baseline load is depicted by a black curve, unmanaged EV loads are depicted in shades of red, optimized loads in shades of green, and loads responding to TOU pricing in shades of blue.  Subfigures (d)-(f) depict the empirical range (lightly shaded region), interdecile range (darkly shaded region), and median (solid line) associated with the aggregate EV load data under each scenario during  Phase  II of the Optimize Project.}
	 \label{fig:dayLoadsConf}
\end{figure}

\section{Flattening peak demand with optimized charging}\label{sec:comparativeEvaluation}
During Phase II of the OptimizEV Project (March 8, 2020 to May 31, 2021), we offered all customers flexibility-differentiated pricing to incentivize delayed charging of their EVs. Utilizing the flexibility provided by customers, a smart charging system continuously optimizes customers' EV charging profiles to flatten the resulting aggregate load profile, while simultaneously ensuring that all customers' EVs are charged by their requested deadlines.  Here, we describe the transformation of aggregate load patterns enabled by this flexibility-differentiated pricing and control mechanism.

As a basis for comparison, we simulate unmanaged EV charging during Phase II of the project. We do so 
by generating an unmanaged charging profile for every charging session initiated during that time frame, where every EV is assumed to draw power at its maximum rate upon connecting to its charging station, until the energy requirement associated with its charging session is fulfilled. Figures \ref{fig:weekLoadUnc} and \ref{fig:weekLoadOpt} depict the aggregate load profiles induced by unmanaged charging and OptimizEV, respectively, between March 9 and March 14, 2020. Notice that unmanaged charging results in a substantial amplification of peak load on these days, while OptimizEV effectively redistributes the aggregate EV load to fill the nighttime valley in the baseline load profile.

To better illustrate the contribution of individual EVs to the aggregate load profile, we disaggregate the EV load profile  across different vehicles under unmanaged and optimized charging in Figures \ref{fig:dayLoadsUnc} and \ref{fig:dayLoadsOpt}, respectively, on March 9, 2020. From Figure \ref{fig:dayLoadsOpt}, it can be seen that EVs participating in optimized charging sessions are typically delayed in time and are charged at lower power levels over longer stretches of time, as compared to their unmanaged counterparts depicted in Figure  \ref{fig:dayLoadsUnc}. It's also worth noting that on this particular day, OptimizEV does not entirely eliminate the contribution of EVs to peak load, as a small subset of users do opt out of optimized charging, deciding instead to have their EVs charged at maximum power without delay. Interestingly, we find that participation rates in optimized charging sessions are strongly correlated with the underlying flexibility (slack) in users' charging requirements, where users are more likely to opt out if they have little flexibility to offer. We discuss user participation rates and trends at  more length in Section  \ref{sec:deadlines}.

Beyond these specific days, the OptimizEV mechanism is also shown to reliably flatten the  aggregate load profile on a majority of days over the fifteen month course of Phase II. To show this, we plot the empirical range, interdecile range, and median associated with the aggregate EV load data  under unmanaged charging  and OptimizEV during Phase II in Figures \ref{fig:dailyUncConf} and \ref{fig:dailyRealizedConf}, respectively.  
As can be seen from these plots, unmanaged charging results in a median increase in peak demand of 6\% (over the baseline peak of 120 kW), while OptimizEV results in a 0\% median increase in peak demand. 
This contrast between unmanaged and optimized charging is even more pronounced when considering days in the bottom 90\% of the aggregate load distribution, where unmanaged charging is shown to induce aggregate load profiles that increase peak demand by as much as 18\%, while OptimizEV only results in a 6\% increase in peak demand on these days.

\begin{figure}[h!]
	\centering
	\includegraphics[width=.65\linewidth, trim=0cm 0cm 0cm 0cm, clip]{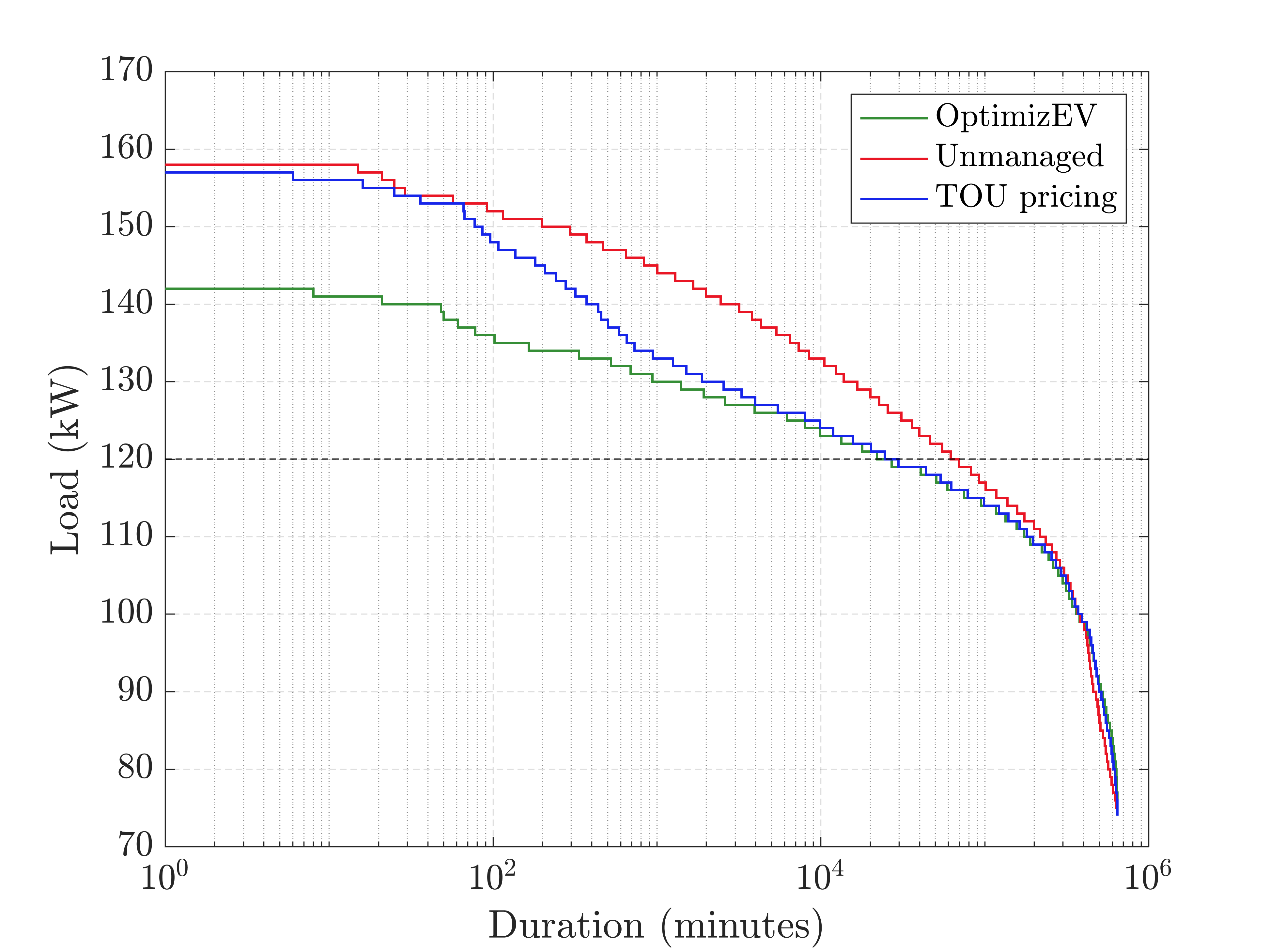}
	\caption{Load duration curves for three scenarios: unmanaged charging (red), OptimizEV (green), and TOU pricing (blue). Each colored curve depicts the number of minutes during which load was greater than or equal to a particular power level throughout Phase II of the OptimizEV Project under the corresponding scenario.
	}
	\label{fig:load_duration}
\end{figure}

We also find that unmanaged charging results in aggregate loads which exceed the baseline peak for significantly longer periods of time than under OptimizEV. In particular, the load duration curves depicted in Figure \ref{fig:load_duration} show that unmanaged charging induces aggregate loads that exceed the baseline peak by at least 20 kW for a total duration of approximately 40 hours over the course of Phase II. In comparison, under OptimizEV, aggregate loads exceed the baseline peak by at least 20 kW for only 20 minutes over that same fifteen-month time span. These findings confirm  that unmanaged charging will likely result in considerable and sustained increases in peak demand, while direct control mechanisms that effectively harness the underlying flexibility in customers' charging requirements can significantly minimize these impacts.

\clearpage

\section{Unintended consequences of time-of-use pricing}
Many electric power utilities and regulators have advocated for the use of nondiscriminatory time-of-use (TOU) electricity rates to minimize the impact of EV charging on peak demand. Under TOU rates, customers are exposed to a high electricity price during ``on-peak" time periods, and a lower electricity price during the complementary ``off-peak" time periods.
TOU rates are designed with the intent of shifting a fraction of the on-peak electricity demand to off-peak periods to flatten the aggregate demand profile. While this might seem like a reasonable approach, the user charging data collected over the course of this project reveals that TOU rates can have perverse consequences.

To understand the potential impact of TOU rates on aggregate demand, we simulate EV charging patterns under  TOU rates using actual customer participation and charging data collected during Phase II of the OptimizEV Project.  For every managed (opt-in) charging session initiated during that time frame, we generate a corresponding delayed charging profile, where a user's EV does not begin charging until the start of the ensuing off-peak pricing period at midnight, at which time the EV begins drawing power at its maximum rate until its energy requirement is satisfied. A charging profile is only delayed to the off-peak period if it is feasible to do so given its underlying energy and completion time requirements. Finally, all of the charging profiles associated with unmanaged (opt-out) charging sessions are left unchanged.

Figure \ref{fig:dailyTOUConf} depicts the empirical median, interdecile range, and range of the aggregate load data simulated under TOU pricing during Phase II of the OptimizEV Project. Notice that TOU pricing is successful in keeping the median aggregate load profile beneath the baseline peak of 120 kW, suggesting that TOU rates may be effective in  combating the adverse effects of unmanaged charging at modest EV penetration levels. However, the synchronization of EV power demand that results when multiple EVs begin charging simultaneously at the start of the off-peak pricing period (midnight) induces new aggregate demand spikes that are sharper and sometimes larger in magnitude than the unmanaged aggregate demand peaks that would have otherwise resulted in the absence of TOU rate-based incentives. This unintended consequence of TOU pricing is clearly  illustrated in Figure \ref{fig:weekLoadTOU}, which depicts a realization of the aggregate load profile simulated under TOU pricing between March 9 and March 14, 2020. Notice that on several days during that week, the peak demand under TOU pricing  significantly exceeds the peak demand realized under unmanaged charging on the same days. These observations suggest that at increased EV penetration levels, passive load coordination mechanisms like TOU pricing will likely fail to attenuate the peak demand driven by EV charging, and may make matters worse.

\section{Why do customers opt out of managed charging?} \label{sec:opt-in}
\label{sec:deadlines}
We find that a large majority of ``opt-out'' sessions are  characterized by inflexible charging requirements. Unsurprisingly, customers were more likely to opt out of managed charging if they  needed their EVs charged very quickly, in time for their upcoming trips.
On the other hand, customers were more inclined to opt in to managed charging if they did in fact possess a reasonable degree of flexibility (slack) in their charging requirements. 

\begin{figure}[htb!]
			\centering			
			\includegraphics[width=.5\linewidth]{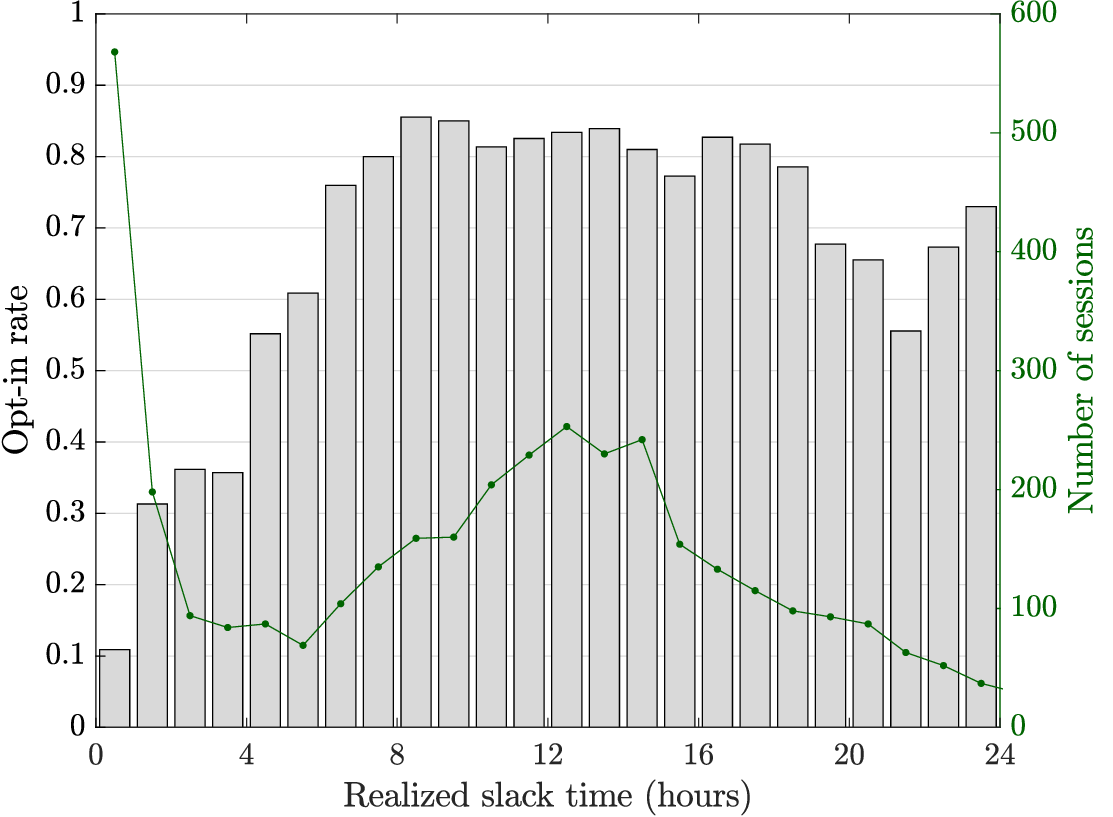}
         \caption{Empirical opt-in frequency as a function of realized session slack time based on all Phase II sessions with realized slack times that are less than or equal to 24 hours. The opt-in frequency associated with Phase II sessions excluded from this plot (i.e., those having realized slack times in excess of 24 hours) is 71\%.}
			\label{fig:slackOptIn}
	\end{figure}

To illustrate this behavior, we plot the empirical opt-in rate as a function of realized session slack time in Figure \ref{fig:slackOptIn}. The \emph{realized slack time} associated with a session is defined as the difference between the actual session duration and the minimum amount of time required to deliver the energy consumed during that session given the user's maximum charging rate. 
Among all Phase II charging sessions, the empirical opt-in rate is lowest (at roughly 10\%) for sessions with realized slack times that are less than one hour. This is unsurprising, since customers with limited flexibility in their charging requirements have very little to offer (or gain) by relinquishing control of their EVs for managed charging. Remarkably, however, the empirical opt-in rate increases steadily as realized session slack times increase from zero to seven hours, and remains relatively constant (hovering around 80\%) for realized session slack times between seven and eighteen hours---revealing that customers are unlikely to opt out of managed charging if they have several hours or more of slack time in their charging requirements.
Surprisingly, the empirical opt-in rate decreases slightly for realized session slack times between 18 and 24 hours. A possible explanation for this behavior is that sessions with very long sojourn times may be associated with lower opt-in rates because they tend to occur on weekends, when customers' commuting patterns are more variable and departure times may be more difficult to anticipate, resulting in a lower willingness to participate in managed charging sessions.
However, because of the relative infrequency of sessions with large slack times,  it is difficult to determine a precise explanation for this observed behavior.

\begin{figure}[htb!]
		\centering
\hspace{1.12in}\includegraphics[width=.7\linewidth]{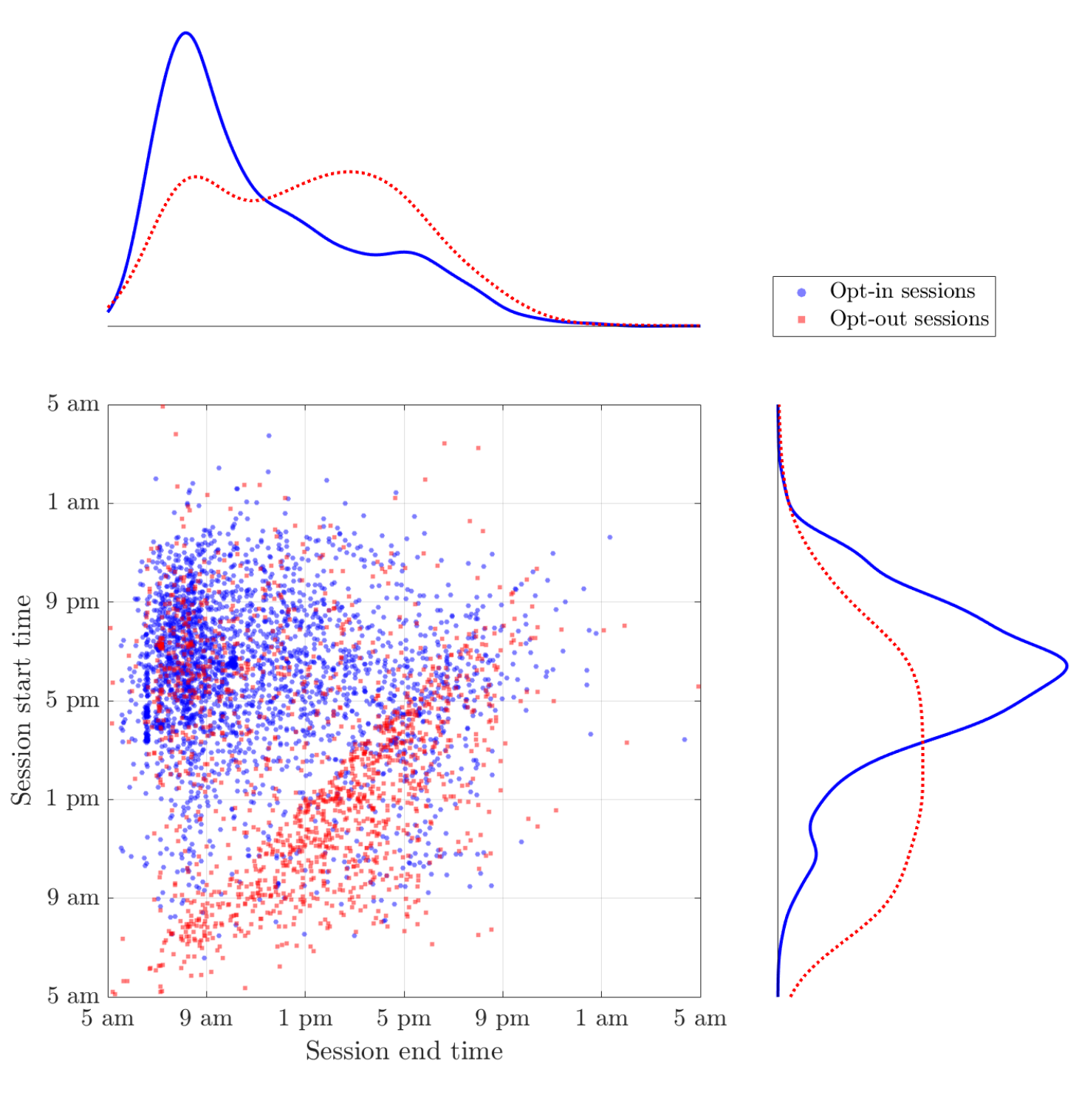}
	\caption{Scatter plot of Phase II session start times versus session end times, where opt-in (opt-out) sessions as blue circles (red squares). The distributions along the vertical and horizontal axes depict  kernel density estimates of the marginal distributions 
	for session start and end times, respectively.}
			\label{fig:slackStart}
\end{figure}

Opt-in and opt-out charging sessions also  differ significantly in terms of their timing. In Figure \ref{fig:slackStart}, we visualize the relationship between Phase II session start times and session end times in the form of a scatter plot that depicts opt-in (opt-out) sessions as blue circles (red squares). Notice that opt-in sessions typically correspond to overnight charging patterns, starting in the evening and ending the following morning. In comparison, opt-out sessions are more commonly associated with daytime commuting patterns, which result in charging sessions that are much shorter in duration and  have start times that are more evenly distributed throughout the day.
This contrast between the timing of opt-in and opt-out sessions suggests that customers are more likely to opt in to managed charging when their commuting needs are flexible and follow regular and predictable patterns, e.g., when they arrive home in the evening and expect to depart the following morning. It is also worth noting that, although opt-out sessions do occur regularly, the uniform dispersion of opt-out session start times throughout the daytime softens their expected contribution to peak demand, as they are unlikely to cluster during on-peak hours.

\ 

\begin{figure}[h!]
	\centering
    \includegraphics[width=1\linewidth]{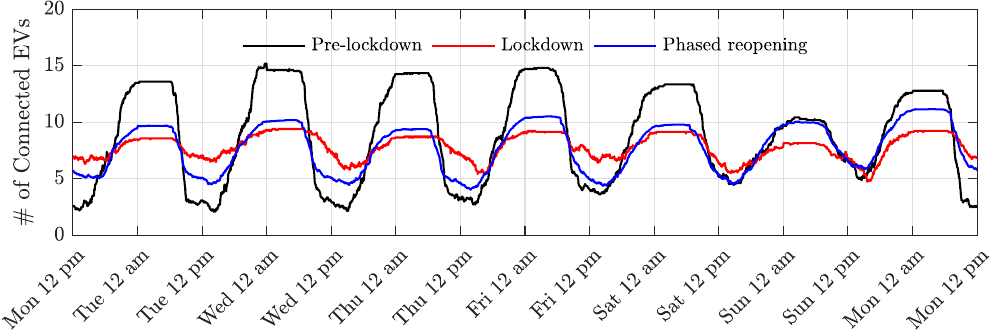}   
	\caption{The plot depicts the average number of EVs connected to the grid throughout the week for three different phases of the COVID-19 pandemic in New York State during this pilot study:   \emph{Pre-lockdown} from January 1, 2020 to March 14, 2020 (black curve),  \emph{lockdown} from March 15, 2020 to June 7, 2020 (red curve), and \emph{phased re-opening} from June 8, 2020 to May 31, 2021 (blue curve).}
	\label{fig:commutingPrePostPan}
\end{figure}
\vspace{-.2in}
\section{COVID-19 impacts on user charging behaviors}
The first recorded U.S. case of COVID-19 was  reported on January 20, 2020. 
In the months following, federal, state, and local governments enacted a sequence of interventions designed to mitigate the spread of infections through ``shelter-in-place" orders and closure of non-essential businesses. Unsurprisingly, the impacts of these interventions can be clearly seen in the shifting charging patterns throughout the OptimizEV Project. In Figure \ref{fig:commutingPrePostPan}, we plot the average number of EVs connected to the grid throughout the week for three different phases of the COVID-19 pandemic. Prior to the lockdown, customers typically connected their vehicles to charge in the evening and disconnected their vehicles the following morning.
During the lockdown, as customers started working from home, their charging patterns---previously dictated primarily by their commute to and from work---became less rigid. 
The average number of vehicles connected throughout the week flattened considerably as customer session start and end times became less concentrated, their session duration became longer, and their charging more infrequent. 
As New York State gradually reopened, user connection patterns partially reverted to pre-lockdown patterns, as charging sessions shortened  and concentrated more  heavily during overnight hours.

\begin{figure}[h!]
	\centering
		\includegraphics[width=1\linewidth, trim  = {1.5cm 4.5cm 1cm 5cm},clip]{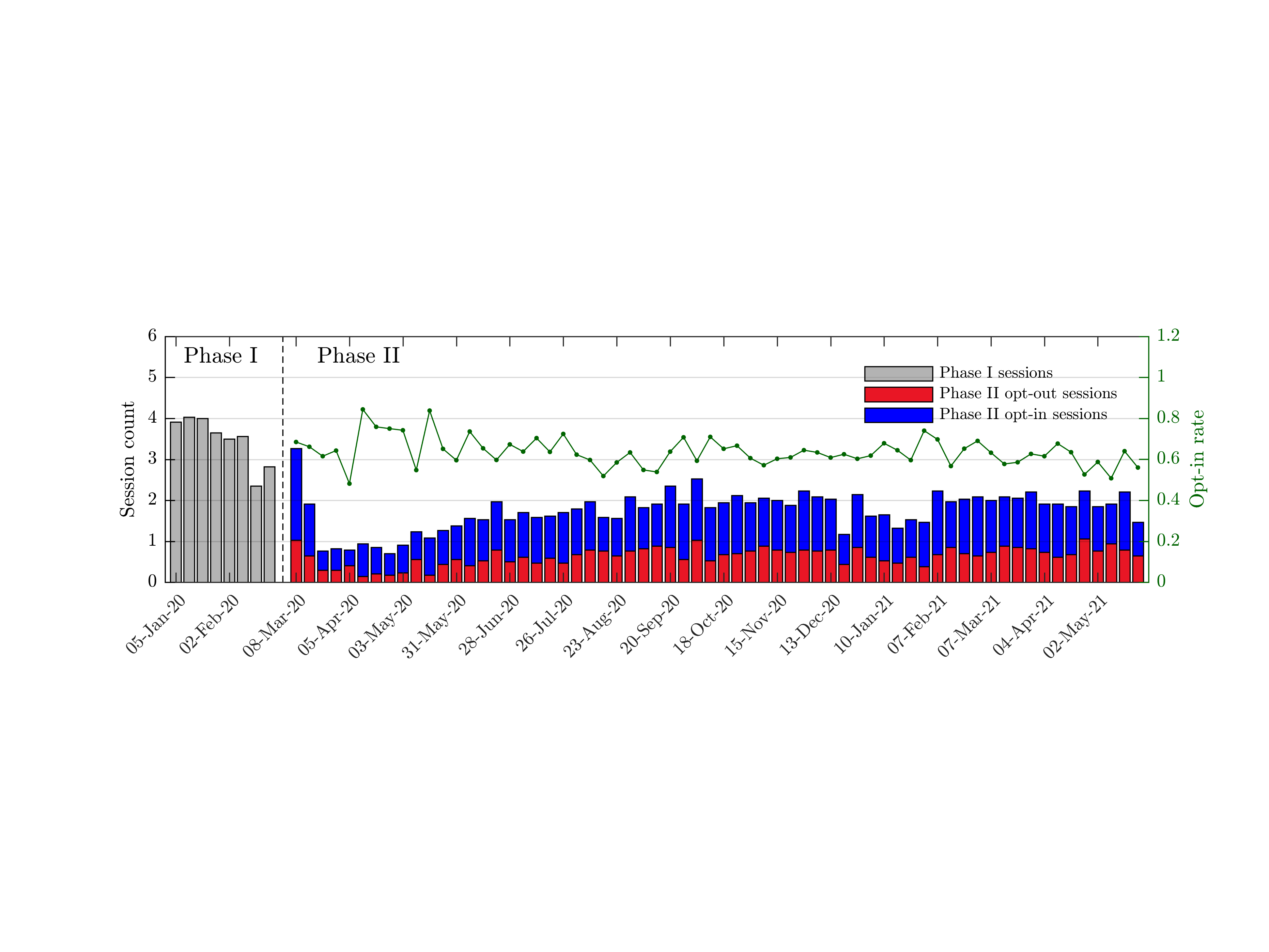}
	\caption{Plot of average number of charging sessions per customer for each week during Phase I (Jan. 1, 2020 to Feb. 29, 2020) and Phase II (Mar. 8, 2020 to May 31, 2021) of the OptimizEV Project (excluding the transition week between Phase I and Phase II). During Phase II, opt-in (opt-out) sessions are depicted in blue (red). The weekly empirical opt-in rate during Phase II is depicted as a green curve. Sessions with zero energy consumption are excluded from these calculations.}
	\label{fig:timeTrends}
\end{figure}

The frequency with which customers charge their EVs also appears to have been impacted by the COVID-19 pandemic. 
Figure \ref{fig:timeTrends} shows the evolution of charging session frequency between January 2020 and May 2021. Prior to the lockdown, customers were found to charge their EVs three or more times per week on average. Immediately after the start  of the lockdown, the average number of weekly charging sessions per customer drops  to approximately 1 session per week, as customers traveled less frequently and spent more time at home.
Between May and November, the average number of weekly charging sessions per customer gradually rises as pandemic related restrictions are loosened.
In December, weekly charging session frequency falls again, likely driven in part by the resurgence in COVID-19 cases during that time frame.  
Despite these shifts in user connection patterns and charging session frequency, the rate at which customers participated in managed charging sessions remained relatively constant over time, hovering around a 60\% opt-in rate throughout Phase II of the pilot study. 
Interestingly, this suggests that customers' willingness to participate in managed charging sessions was not impacted by the various lockdown and reopening measures implemented over the course of the COVID-19 pandemic.

\vspace{-.2in}
\section{Conclusion} \label{sec:conclusion}
The  flexibility-differentiated pricing and control mechanism studied as part of the OptimizEV Project was shown to be highly effective  in reshaping residential EV charging loads to minimize their impact on peak electricity demand. The observed effectiveness of the mechanism is due in large part to high customer participation rates---when customers exhibited flexibility in their charging requirements, they frequently engaged in managed charging sessions, allowing the smart charging system to delay the completion of their charging by nine hours on average. 
 
While these initial observations provide concrete evidence in support of the proposed incentive and control mechanism as a potentially viable ``non-wires alternative'' to support the increased demand for electricity driven by the growing adoption of EVs, larger experimental studies and surveys are needed to better understand customer attitudes towards managed charging and the feasibility of such programs at scale.
 In particular, because of the differences between early EV adopters and the general population of electricity customers \citep{powells2019flexibility,fjellsaa2021justice}, studies involving a more diverse population of customers are necessary.
 Although personal vehicle usage patterns in the general population suggest the presence of significant charging flexibility  across the general population \citep{anwar2022assessing}, it is difficult to ascertain how a typical person's willingness to engage in managed charging compares to that of the project participants.  
Moreover, because the typical electricity customer tends to be less technologically savvy and less motivated by environmental concerns  than the early EV adopters participating in the OptimizEV Project, the monetary incentive required to elicit flexibility from the general population may be larger.
 
Finally, in order for managed EV charging programs like OptimizEV to be cost effective at scale, the benefit to the utility company, in the form of reduced energy procurement or avoided infrastructure costs, must exceed the program implementation and operational costs. 
These include administrative costs, the total cost of incentives paid out to participating customers, and the cost of the additional metering, communication, and control equipment required to directly manage EV charging processes. 
If the estimated benefits are expected to exceed the costs, then it may be prudent for the utility company to pursue programs of this kind to manage the increasing penetration of EV loads on their networks.
\section{Acknowledgments}
There are a number of partners without whom the OptimizEV Project would not have been possible. These partners include New York State Electric and Gas (NYSEG),  New York State Energy Research and Development Authority (NYSERDA),  New York State Department of Public Service, Cornell Cooperative Extension,  Taitem Engineering, and Kitu Systems. This research was supported in part by the the Holland Sustainability Project Trust and the  National Science Foundation under a Graduate Research Fellowship, Grant No. ECCS-135162, and Grant No. TPI/TI-1632124.

\appendixtitleon
\begin{appendices}

\section{Methods} \label{sec:methods}

\subsection{Baseline non-EV load profile}\label{sec:baseline}

Throughout Phase II of the OptimizEV Project, we utilized a fixed baseline non-EV load profile as an input to our real-time scheduling algorithm. The load profile is based on a 24-hour long time series of aggregate load measurements taken from a distribution circuit in Tompkins County, New York, which serves 623 residential customers and 69 commercial and industrial customers. 
In order to obtain a load profile that reflects a small residential neighborhood consisting of 60 households, we re-scaled the given time series so that the resulting peak load equals 120 kW.

\subsection{Real-time scheduling algorithm} \label{sec:schDetail}

In what follows, we provide a detailed mathematical formulation of the real-time scheduling algorithm that was used  as part of the OptimizEV  Project. We index time periods by $t=1,2, \dots$, and let $\Delta = (1/60)$ (units: hours) denote the length of each  time period. 
We let  $\mathcal{N}(t) \subseteq \{1, \dots, 34\}$ denote  the set of EVs that are actively engaged in \emph{managed} (opt-in) charging sessions at time $t$. 
The charging profiles associated with EVs engaged in managed sessions can be adjusted within the constraints implied by their users' stated charging requirements. We denote the charging requirements associated with each user $i \in \mathcal{N}(t)$ at time $t$ by
\begin{itemize}
    \item $R_i$:  user $i$'s maximum charging rate (units: kilowatts),
    \item $E_i(t)$: user $i$'s residual energy requirement at time $t$ (units: kilowatt-hours),
    \item $d_i(t)$: number of remaining time periods  by which the energy requirement  $E_i(t)$ must be satisfied.
\end{itemize}
The SAE J1772 charging protocol also imposes a charging constraint in the form of a minimum non-zero charging rate, which we denote by $\minRate$.
At each time $t$, the scheduling algorithm generates a charging profile for every actively managed EV spanning the next $T = 1440$ time periods (24 hours). We denote the sequence of charging power commands sent to each  active EV $i \in \mathcal{N}(t)$ at time $t$ by  $r_i(1), \, r_i(2), \, \dots, \, r_i(T) \ \text{(units: kilowatts).}$
The managed  charging profiles are chosen to collectively flatten the aggregate load profile by solving the following optimization problem at each time $t$:  
\begin{linenomath*}
\begin{subequations}
	\label{prob:opt}
	\begin{alignat}{3}
	& \text{minimize}  \quad & &  \sum_{k=1}^{T} \left(  \baseload(t+k-1) +  \sum_{i \in \activeCusts}   \decVarCustTime   \right)^2  \label{eqn:objective} \\
	& \text{subject to} & & \sum_{k=1}^{d_i \left(t\right)} \decVarCustTime  \Delta = \energyResCust, \hspace{.41in} \forall \,  i \in \activeCusts, \label{prob:opt_energy}  \\
	& & & \decVarCustTime \in \left\{ 0  \right\} \cup \left[\minRate, \, \maxRate_i \right],  \hspace{.11in} \forall \,  k \in \left\{1, \dots, d_i(t) \right\} \ \, \text{and} \ \, i \in \activeCusts,    \label{prob:opt_rate} \\
	& & & r_i(k) = 0, \hspace{1.04in} \,   \forall \,  k \notin \left\{1, \dots, d_i(t) \right\}  \ \, \text{and} \ \,  i \in \activeCusts. \label{prob:nocharge}
	\end{alignat}
\end{subequations}
\end{linenomath*}
In the above optimization problem, the sequence $\ell(t),  \, \dots, \; \ell(t+T-1)$ (units: kilowatts) denotes the baseline non-EV load profile plus the aggregate \emph{unmanaged} EV load profile stemming from all EVs engaged in unmanaged  (opt-out) charging sessions at time $t$. Note that  EVs engaged in unmanaged charging sessions are charged at their maximum rates, without delay, until their energy requirements are satisfied. Constraint \eqref{prob:opt_energy} ensures that each user's requested energy is delivered  prior to their deadline.
Constraint \eqref{prob:opt_rate} ensures that the charging power commands computed for each actively managed EV are either zero valued, or between the charging station's minimum non-zero charging rate and  the EV's maximum charging rate. Constraint \eqref{prob:nocharge} ensures that no actively managed EV is charged after its deadline has passed. The optimization criterion, defined on line \eqref{eqn:objective}, promotes EV charging profiles with a ``valley-filling'' characteristic.

In  order  to  effectively  adapt  to  changing  system  conditions  in  real  time,  the  smart-charging  system developed  for  the  OptimizEV  Project  utilizes  a  model  predictive  control  approach  to  continuously re-optimize  the  active  EVs’  charging  profiles  every  minute  of the  day.  We provide pseudocode that clearly specifies the recursive nature of this real-time optimization routine in Algorithm \ref{alg:mpc}. 

\begingroup
\nolinenumbers

 \

\begin{algorithm}[H]
	\For{$t \in \left\{1,  2, \dots \right\}$ } {
	Solve optimization problem \eqref{prob:opt} to compute optimal charging profile $(r^{*}_i(1), \dots r^*_i(T))$ for each actively managed EV $i \in  \activeCusts$ \;
\For{$ i \in \activeCusts$ } {
	            Transmit optimal charge rate ${r}^{*}_{i}(1) $ to EV $i$ \;
				Measure charge rate $\widehat{r}_i \left(t \right)$ implemented by EV $i$ \;
				Set $E_i \left( t+1 \right) := \energyResCust - \widehat{r}_i \left(t\right) \Delta$ \;
				Set $d_i \left(t+1 \right) := d_i \left(t \right) - 1$ \;
			}
	}
	\caption{Real-time Scheduling Algorithm}
	\label{alg:mpc}
\end{algorithm}

\

\endgroup

We briefly summarize each step of the pseudocode provided in Algorithm \ref{alg:mpc}. Immediately prior to time period $t$, the smart-charging system solves problem \eqref{prob:opt} to determine optimal charging profiles for all actively managed EVs, which we denote by $r^{*} = \left\{ r^{*}_i\left( k \right) \mid i \in \activeCusts, k \in \{1, \dots, T\}  \right\}$.  These optimal charging profiles are then transmitted to each EV's charging station as a sequence of time-varying power commands. Each EV $i \in \mathcal{N}(t)$ is instructed to execute the first component of the sequence of power commands that it receives,  $r^{*}_i(1)$. Because there may be deviations between the actual power drawn by an EV and the commanded power, each active EV $i$ transmits a measurement of the actual power drawn during time period $t$ to the smart-charging system, which we denote by $\widehat{r}_i(t)$. The smart-charging system uses this information to update   each user's residual energy requirement according to $E_i(t+1) = E_i(t) -  \widehat{r}_i(t) \Delta$. This information, together with an updated forecast of the baseline non-EV load and aggregate unmanaged EV load, is used to resolve problem \eqref{prob:opt}, generating a new sequence of power commands for the subsequent time period $t+1$.

\subsection{Practical algorithmic considerations} \label{sec:practical}
In this section, we discuss a number of practical challenges encountered when deploying the proposed real-time scheduling algorithm in a real-world setting.  This first challenge pertains to the need to ensure that all necessary computations that are carried out at each time period  are completed  within the allowable time budget (less than one minute).  Problem \eqref{prob:opt} is a nonconvex optimization problem due to the disjunctive constraint \eqref{prob:opt_rate}.
Specifically, it belongs to the class of mixed integer quadratic programs (MIQP).
Although it is possible to solve MIQPs exactly using branch and bound methods, these approaches can be extremely slow when the number of decision variables is high. To circumvent these computational challenges, we employ a convex relaxation of the disjunctive constraint \eqref{prob:opt_rate} by allowing the charging rate to take any value between zero and the maximum charging rate. The resulting relaxation is a convex quadratic program that can be solved to optimality within seconds on a standard desktop computer using interior point methods. We convert solutions generated by the relaxed problem into feasible solutions for the original problem by appropriately rounding those solutions to ensure that the original disjunctive constraint \eqref{prob:opt_rate} is satisfied. 

The second challenge is related to the occasional constraint infeasibility that results from  nonidealities in EV charging characteristics and uncertainty in user inputs. Specifically, in order to maintain feasibility of the optimization problem \eqref{prob:opt}, it must be possible to satisfy every user's residual energy requirement prior to their stated deadline subject to their maximum rated charging capacity. In other words, the slack associated with every active charging requirement must always be non-negative. Formally, this corresponds to the requirement that 
\begin{linenomath*}
	\begin{align}
	\deadlineCustTime  - \frac{\energyReq_i(t)}{  \maxRate_i \Delta }  \, \geq  \, 0,
	\label{eqn:feas}
	\end{align}
\end{linenomath*}
for all users $i \in \mathcal{N}(t)$ and time periods $t$.
In order to enforce the satisfaction of this requirement, users are not permitted to make infeasible charging requests. However, because the power drawn by an EV may deviate from the power commands that it's instructed to follow, charging requests may become infeasible over time. When a user's charging request becomes infeasible, it is modified by reducing the user's residual energy requirement $\energyReq_i(t)$ to the maximum amount of energy that can be delivered prior to the user's stated deadline, to ensure the feasibility of the optimization problem  \eqref{prob:opt}.

\subsection{Demographics} \label{sec:demo}

While demographic data on the participants included in the survey is not available, a study of recipients of a rebate for new electric vehicles by the New York State Energy Research and Development Authority (NYSERDA) provides insight into the demographics of EV drivers in New York State \citep{nyserda2021consumer}. 
Study respondents were educated, with 77\% having completed a Bachelor's degree. 
Respondents tended to be well-off, with 89\% reporting that they owned their home and 67\% having an annual household income of at least 100,000 USD.
Additionally, respondents demonstrated significant interest in sustainability: 82\% of respondents reported that reducing energy impacts was of high importance in their choice to purchase an EV.

\end{appendices}

\bibliographystyle{agsm}
\bibliography{References}

\end{document}